\documentclass[11pt]{article}
\usepackage{amsfonts}
\usepackage{amssymb}
\usepackage{amsmath}
\usepackage{amstext}
\usepackage{trackchanges}

\renewcommand{\thepage}{}

\addeditor{HI}
\addeditor{WN}
\input{tcilatex}
\begin{document}

\title{The Influence Function of Semiparametric Estimators\thanks{%
The JSPS 15H05692 and 20H00072 and NSF Grants SES 1132399 and 1757140
provided financial support. We are grateful for comments by X. Chen, V.
Chernozhukov, D. Hughes, K. Kato, R. Matzkin, U. M\"{u}ller, Y. Mukhin, J.
Porter, D. Pouzo, A. Santos and participants at seminars at UC Berkeley,
NYU, University of Kansas, and Yale. L. Hoderlein provided capable research
assistance.}}
\author{Hidehiko Ichimura \\
University of Arizona and University of Tokyo \and Whitney K. Newey \\
MIT}
\date{July 2015\\
Revised April 2021}
\maketitle

\begin{abstract}
There are many economic parameters that depend on nonparametric first steps.
Examples include games, dynamic discrete choice, average exact consumer
surplus, and treatment effects. Often estimators of these parameters are
asymptotically equivalent to a sample average of an object referred to as
the influence function. The influence function is useful in local policy
analysis, in evaluating local sensitivity of estimators, and constructing
debiased machine learning estimators. We show that the influence function is
a Gateaux derivative with respect to a smooth deviation evaluated at a point
mass. This result generalizes the classic Von Mises (1947) and Hampel (1974)
calculation to estimators that depend on smooth nonparametric first steps.
We give explicit influence functions for first steps that satisfy exogenous
or endogenous orthogonality conditions. We use these results to generalize
the omitted variable bias formula for regression to policy analysis for and
sensitivity to structural changes. We apply this analysis and find no
sensitivity to endogeneity of average equivalent variation estimates in a
gasoline demand application.
\end{abstract}

\bigskip

\bigskip

\bigskip

\bigskip

\bigskip

\bigskip

\bigskip

\bigskip

\bigskip

\bigskip

\bigskip

\bigskip

\bigskip

\bigskip



\baselineskip=22pt\newpage \setcounter{page}{1}\renewcommand{\thepage}{[%
\arabic{page}]}\renewcommand{\theequation}{\arabic{section}.%
\arabic{equation}} \setlength{\baselineskip}{18pt}

\section{Introduction}

There are many estimators of economic parameters that depend on
nonparametric first steps. Examples include games, dynamic discrete choice,
average consumer surplus, and treatment effects. Often these estimators are
asymptotically equivalent to a sample average. The object being averaged is
referred to as the influence function.

The influence function has several important uses. It can be used for
quantifying local policy effects. For example, Firpo, Fortin, and Lemieux
(2009) use influence functions to quantify local policy effects of changes
in explanatory variables on quantiles or other characteristics of a
distribution. We give local policy effects of structural changes. The
influence function can also be used to measure sensitivity of estimators to
misspecification. Its use for qualitative sensitivity measures is where the
influence function gets its name in the robust estimation literature, see
Hampel (1974). The expected GMM influence function under a misspecified
distribution is the GMM sensitivity measure given in Andrews, Gentkow, and
Shapiro (2017). We quantify sensitivity for objects that depend on solutions
to orthogonality conditions. We use this quantification to generalize the
classic omitted variables bias formula for regression coefficients to many
other objects. We apply these results to estimate sensitivity of equivalent
variation bounds to endogeneity of gasoline demand.

Another important use of the influence function is construction of
orthogonal moment functions where first step estimation has no first order
effect on moments. Orthogonal moment functions reduce bias in GMM from model
selection and regularization of the first step and enable machine learning
for high dimensional first steps, as in Chernozhukov et al. (2018) and
Chernozhukov et al. (2020). The influence function formulae given here are
used in Chernozhukov et al. (2020) to derive orthogonal moment functions.
The influence function can also be used to compare asymptotic efficiency of
estimators and find efficient ones. Efficient estimation is important in
many econometric settings where weak assumptions are made to make models
empirically plausible. Knowing the form of the influence function also
facilitates asymptotic theory by showing in advance the conclusion of an
asymptotic expansion.

Newey (1994) showed that the influence function of an estimator could be
obtained from the probability limit (plim) of the estimator. A functional
equation was given that can be solved for the influence function without an
asymptotic, large sample expansion. Hahn (1998) and Hirano, Imbens, and
Ridder (2003) applied this approach to derive the influence function of
important treatment effect estimators. A primary purpose of this paper is to
give a simpler way of calculating the influence function and to illustrate
its usefulness for applied researchers. We show that the influence function
can be calculated from a derivative of the plim with respect to a scalar
mixture of the true distribution with another distribution. This calculation
extends the classic Von Mises (1947), Hampel (1974), and Huber (1981)
Gateaux derivative calculation to objects that exist only for continuous
distributions. We also illustrate how this Gateaux derivative can be used to
facilitate empirical research on local policy analysis, quantify sensitivity
of estimators, and construct orthogonal moment functions.

The functional equation in Newey (1994) has been solved to obtain influence
functions in many important settings. Newey (1994) did so for estimators
that depend on a first step least squares projection or a probability
density function (pdf). Bajari, Hong, Krainer, and Nekipelov (2010) and
Bajari, Chernozhukov, Hong, and Nekipelov (2009) did so for game models and
Hahn and Ridder (2013, 2016) did so for nonparametric generated regressors.
We use the Gateaux derivative calculation to derive influence functions for
first steps that solve orthogonality conditions, both exogenous and
endogenous. These calculations provide explicit influence function formulae
for a variety of estimators in addition to those already in the literature.
The calculations also illustrate the simplicity and usefulness of the
formulae here in making the influence function more widely available for
empirical research involving local policy analysis, estimator sensitivity,
or orthogonal moment functions.

We estimate sensitivity to endogeneity of bounds on average equivalent
variation for gasoline demand. This application is motivated by the
difficulty of simultaneously allowing for price endogeneity and general
preferences in demand analysis. Hausman and Newey (2016) gave nonparametric
estimators of bounds on average equivalent variation with general
preferences that are independent of prices and income. For scalar
heterogeneity and endogenous prices Blundell, Horowitz, and Parey (2017)
estimate the gasoline demand function via nonparametric quantile
instrumental variables, which is computationally difficult and only allows
scalar heterogeneity. The bound sensitivity we give is much simpler and
allows general heterogeneity. We find that for gasoline demand the average
equivalent variation bounds are not very sensitive to endogeneity and that
the sensitivity is not statistically significant.

{A distinctive feature of our approach is that the influence function is
obtained directly from the moment conditions defining the estimator without
solving an integral equation or going through a probabilistic calculations
in the form of asymptotic arguments. In this sense, our result allows us to
study semiparametric estimators analogously to the estimators obtained based
on the parametric maximum likelihood or the generalized method of moment
conditions. Using asymptotic arguments} Robinson (1988), Powell, Stock, and
Stoker (1989), Goldstein and Messer (1992), Ichimura (1993), Klein and Spady
(1993), and Chaudari, Doksum, and Samarov (1997) gave influence function
formulae for important semiparametric estimators. Newey (1994) gave general
explicit influence function formulae where a first step is an infinite
dimensional regressions or pdf. Ai and Chen (2007, 2012), Ichimura and Lee
(2010), Ackerberg et al. (2014), Chen and Liao (2015), and Chen and Pouzo
(2015) gave interesting and useful characterizations of influence functions
for estimators with first steps that solve conditional moment restrictions
or that are maximizers of an objective function. The results of this paper
are complementary to this previous work in providing explicit formulae for
influence functions for estimators that solve orthogonality conditions. Such
explicit formula are useful for policy and sensitivity analysis and for
construction of orthogonal moment functions.

A primary objective of this paper is to provide a method to compute the
influence functions for semiparametric estimators. The influence function of
an estimator may be different than the efficient influence function for the
parameter of a semiparametric model considered e.g. by Bickel et al. (1993).
These do coincide in models where a parameter is exactly identified; see
Chen and Santos (2015). One can think of the object derived here as the
efficient influence function for the parameter that is defined as the plim
of an estimator for a general, unrestricted distribution. This parameter is
exactly identified in the model with the unrestricted distribution so the
efficient influence function coincides with the influence function of the
estimator. This is the approach taken by Newey (1994) to finding the
influence function of an estimator. We simplify this approach in a way that
makes it more applicable to empirical research.

Validity of the influence function calculation given here depends on
distributional variation that is a smooth approximation to a distribution
that puts all probability on a point, i.e. is a point mass. After the first
version of this paper appeared on arXiv, Luedtke, Carone, and van der Laan
(2015) and Carone, Luedtke, and van der Laan (2016) used such deviations in
estimation. This construction is useful in that setting, but we emphasize
that we have a different goal here; to calculate the influence function of
any semiparametric estimator.

Muhkin (2019) used the influence function to derive local effects of
changing one object of interest on another object of interest. Also, the
local effects are integrated to obtain global effects. This work also shows
the usefulness of the influence functional calculation given here.

Summarizing, the contributions of this paper are to i) give a simpler way of
calculating the influence function; ii) derive explicit influence function
formulae for functions satisfying exogenous and endogenous orthogonality
conditions; iii) give local policy effects and sensitivity to structural
changes and illustrate their use in empirical research; and iv) show absence
of local sensitivity to endogeneity of equivalent variation in a gasoline
demand application.

In Section 2 we give the Gateaux derivative formula for the influence
function and describe several important uses of this formula. Section 3
gives the influence function for exogenous orthogonality conditions and uses
that to derive local policy effects and sensitivity for structural change.
It is shown that these formula generalize the classic omitted variables bias
formula. Section 4 gives the influence function for endogenous orthogonality
conditions. Section 5 discusses extensions and conclusions. Appendices give
regularity conditions for validity of the influence function calculation,
characterize the influence function for minimum distance estimators, and
extend the explicit influence function formulae to misspecified
orthogonality conditions with endogenity.

\section{The Influence Function and Its Uses}

The estimators and objects in this paper are allowed to depend on a first
step nonparametric estimator. We refer to these estimators as
semiparametric. We denote such an estimator by $\hat{\theta}$, which is a
function of the data $W_{1},...,W_{n}$ where $n$ is the number of
observations. Throughout the paper we will assume that the data observations 
$W_{i}$ are i.i.d. with some cumulative distribution function (CDF) $F_{0}$.
We let $\theta _{0}$ denote the probability limit of $\hat{\theta}$ when $%
F_{0}$ is the distribution of $W_{i}$.

In this paper we focus on asymptotically linear estimators that satisfy%
\begin{equation}
\sqrt{n}(\hat{\theta}-\theta _{0})=\frac{1}{\sqrt{n}}\sum_{i=1}^{n}\psi
(W_{i})+o_{p}(1),\text{ }E[\psi (W)]=0,\text{ }E[\psi (W)^{T}\psi
(W)]<\infty .  \label{inf}
\end{equation}%
The asymptotic variance of $\hat{\theta}$ is then $E[\psi (W)\psi (W)^{T}]$.
The function $\psi (w)$ is referred to as the influence function, following
terminology of Hampel (1974). It gives the influence of a single observation
in the leading term of the expansion in equation (\ref{inf}). It also
quantifies the effect of a small change in the distribution of $W$ on the
probability limit of $\hat{\theta}$ as we further explain below. Very many
root-n consistent semiparametric estimators are asymptotically linear under
sufficient regularity conditions, including M-estimators, Z-estimators,
estimators based on U-statistics, and many others; see Bickel, Klaasen,
Ritov, and Wellner (1993) and Van der Vaart (1998).

The influence function of an estimator can be obtained without deriving the
stochastic expansion in equation (\ref{inf}) as was in Newey (1994). Let $F$
be any distribution that is unrestricted except for regularity conditions
and $\theta (F)$ denote the probability limit of $\hat{\theta}$ when $F$ is
the CDF of $W$. Here $\theta (F)$ can be thought of as the probability limit
of $\hat{\theta}$ under general misspecification where $F$ is only required
to satisfy some regularity conditions (like some random variables being
continuously distributed and/or existence of certain moments) but is
otherwise unrestricted. Also, let $\{F_{\beta }\}$ be any parametric family
of distributions passing through $F_{0}$ with $F_{\beta }=F_{0}$ when $\beta
=0$ and satisfying certain regularity conditions with score (derivative of
log-likelihood) $S_{\beta }(w)$ at $\beta =0.$ Then by Van der Vaart (1991)
it follows that the influence function satisfies%
\begin{equation}
\frac{\partial \theta (F_{\beta })}{\partial \beta }=E[\psi (W)S_{\beta
}(W)],  \label{score formula}
\end{equation}%
when the estimator $\hat{\theta}$ is locally regular in the sense discussed
in Van der Vaart (1991). This is a functional equation from which $\psi (w)$
may be obtained by varying $\{F_{\beta }\}$ and the associated score $%
S_{\beta }(w)$. In several important settings the influence function has
been obtained by solving this functional equation without the stochastic
expansion in equation (\ref{inf}). Newey (1994) did this for first step
regression and density estimation. Hahn (1998) obtained the influence
function for the regression estimator of the average treatment effect and
Hirano, Imbens, and Ridder (2003) for inverse propensity score weighted
estimators. Hahn and Ridder (2013, 2016) did so for first step generated
regressors and control functions and Bajari et al. (2009, 2010) for
estimating game models.

A main purpose of this paper is to give a simpler, more direct way of
calculating the influence than solving equation (\ref{score formula}). Let $%
H $ denote a CDF such that $\theta (F_{\tau })$ exists for $F_{\tau
}=(1-\tau )F_{0}+\tau H$ where $\tau $ is a scalar with $0\leq \tau <C$ for $%
0<C<1.$ Equation (\ref{inf}) and regularity conditions discussed in Appendix
A imply that%
\begin{equation}
\frac{d\theta (F_{\tau })}{d\tau }=\int \psi (w)H(dw),\text{ }E[\psi (W)]=0,%
\text{ }E[\psi (W)^{2}]<\infty ,  \label{G deriv}
\end{equation}%
where throughout the paper $d/d\tau $ denotes a derivative from the right at 
$\tau =0.$ This equation suggests a direct way to calculate the influence
function:

\bigskip

STEP I: Calculate $d\theta (F_{\tau })/d\tau $ for any $H$ such that the
derivative exists;

\bigskip

STEP II: Evaluate the derivative formula at $H=\Delta _{w},$ where $\Delta
_{w}$ is the CDF with $\Pr (W=w)=1,$ to obtain $\psi (w)=\int \psi (\tilde{w}%
)\Delta _{w}(d\tilde{w})$ as a function of $w.$

\bigskip

Equation (\ref{G deriv}) does not justify Step II because the derivative
need not exist when $H=\Delta _{w}$. In particular $\theta (F_{\tau })$ may
not be well defined when $\theta (F)$ depends on a pdf or conditional
expectation because of the discrete component $\Delta _{w}$ of $F_{\tau
}=(1-\tau )F_{0}+\tau \Delta _{w}.$ The nonexistence of a pdf of $F_{\tau }$
of $(1-\tau )F_{0}+\tau \Delta _{w}$ at any $\tau >0$ can make $\theta
(F_{\tau })$ undefined. Nevertheless Step II is justified as a limit as $H$
approaches $\Delta _{w}$, similar to Lebesgue (1904) differentiation in
analysis; e.g. see Wheeden and Zygmund (1977). A precise justification for
Step II is given in Appendix A.

The calculation in Steps I and II generalizes the classic Hampel (1974)
formula, $\psi (w)=d\theta ((1-\tau )F_{0}+\tau \Delta _{w})/d\tau ,$ to
cases where existence of $\theta (F)$ requires some components of $W$ be
continuously distributed. Such cases are very important for semiparametric
estimators where $\theta (F)$ can depend on limits of nonparametric
estimators of densities, conditional expectations, or other objects whose
existence requires $W$ have continuously distributed components. Steps I and
II provide a simpler and more direct way of obtaining $\psi (w)$ than
solving the integral equation (\ref{score formula}).

The influence function does not exist when $\theta (F)$ does not satisfy the
Stein (1956) necessary conditions for existence of a root-n consistence
estimator. In that case Steps I and II will fail. To illustrate, suppose $W$
is continuously distributed with pdf $f_{0}(w)$ and $\theta (F)=f(\bar{w})$
is the pdf of $W$ at some fixed $\bar{w}.$ In that case%
\begin{equation}
\frac{d\theta (F_{\tau })}{d\tau }=h(\bar{w})-f_{0}(\bar{w}).
\label{failure}
\end{equation}%
Because $h(\bar{w})$ is the pdf of $H(w)$ at the point $\bar{w}$ it cannot
be represented as the expectation over $H$ of a function with finite second
moment. In general Steps I and II will fail whenever equation (\ref{G deriv}%
) is not satisfied. As in equation (\ref{failure}), this failure will often
be evident in the calculation of $d\theta (F_{\tau })/d\tau .$

Equation (\ref{G deriv}) motivates the use of the influence function in
empirical work. The Gateaux derivative $d\theta (F_{\tau })/d\tau $ is the
local effect of changing the distribution $F$ on the object $\theta (F).$ If
we broaden the interpretation of $\theta \left( F\right) $ to include
economic objects of interest, such as a feature of the distribution of
outcome variables, then $d\theta (F_{\tau })/d\tau $ can be thought of as a
local policy effect of changing the distribution of the data. Equation (\ref%
{G deriv}) then can be used to obtain the local policy effect from the
influence function, as did Firpo, Fortin, and Lemeiux (2009) for the policy
effect of changing the distribution of regressors. When $\theta (F)$ is the
probability limit of an estimator $\hat{\theta}$ we can think of $d\theta
(F_{\tau })/d\tau $ as the local sensitivity of that estimator to changes in 
$F$, which gives local effects of misspecification. The GMM sensitivity
analysis of Andrews, Gentzkow, and Shapiro (2017) has precisely the form of
equation (\ref{G deriv}), as will be discussed in Section 2.2. In addition,
when $\theta \left( F\right) $ is the true expectation of an identifying
moment function evaluated at the limit of a first step estimator, the
influence function can be used to create orthogonal moments that have zero
Gateaux derivative with respect to the first step. As discussed in
Chernozukov et al. (2018), this use of the influence function is helpful for
debiased machine learning of objects of interest.

In the remainder of this Section we describe more fully these important uses
of the influence function that are of direct interest to empirical
researchers. Here we show how this paper can be applied to obtain novel
policy effects of structural change, local sensitivity measures and Hausman
tests, and orthogonal moment functions.

\subsection{Local Policy Analysis of Structural Changes}

In many settings $\theta (F)$ may be an economic quantity of interest.
Changes in $F$ can sometimes be thought of as changes in a policy. From
equation (\ref{G deriv}) we see that $\int \psi (w)H(dw)$ is the derivative
of $\theta (F)$ as $F_{\tau }$ changes away from $F_{0}$ in the direction $%
H-F_{0}$. If $H$ is thought of as resulting from a change in policy, then $%
\int \psi (w)H(dw)$ will be the derivative of the economic quantity of
interest with respect to that policy change, i.e. a local policy effect.

Firpo, Fortin, and Lemeiux (2009) derive such effects where $\theta (F)$ is
specified as some feature of the marginal distribution of an outcome
variable $Y$ and the change in policy is a change in the distribution of
explanatory variables $X$. Because $\theta (F)$ depends only on the marginal
distribution of $Y,$ the influence function of $\theta (F)$ will be $\psi
(y) $ that depends only on $y$. For example if $\theta (F)$ is the $p^{th}$
quantile of $F,$ satisfying $F_{Y}(\theta (F))=p,$ then $\psi
(y)=[1(y<\theta _{0})-p]/f_{Y0}(\theta _{0}),$ where $f_{Y0}(y)$ is the true
marginal pdf of $Y.$ Because the distribution of $X$ is different in $H$ but
nothing else is different than in $F_{0},$ the conditional distribution of $%
Y $ given $X$ will be the same for $H$ as it is for $F_{0}.$ Then by
iterated expectations the local policy effect is%
\begin{equation*}
\frac{d\theta (F_{\tau })}{d\tau }=\int \psi (y)H(dw)=E_{H}[\psi
(Y)]=E_{H}[E[\psi (Y)|X]].
\end{equation*}%
Firpo, Fortin, and Lemeiux (2009) analyze such policy effects for quantiles
of $Y$, other objects $\theta (F)$ of interest, and for a variety of
alternative policy shifts in the distribution of $X$ as represented by $H.$

One can also specify the policy effect of a structural change where the
conditional distribution of $Y$ given $X$ changes and the marginal
distribution of $X$ remains unchanged. The local policy effect of a
structural change is%
\begin{equation*}
\frac{d\theta (F_{\tau })}{d\tau }=E_{H}[\psi (Y)]=E[E_{H}[\psi (Y)|X]].
\end{equation*}%
Here we see the that local effect of a structural change in the direction $%
H-F_{0}$ is captured by the conditional expectation $E_{H}[\psi (Y)|X]$ of
the influence function $\psi (Y)$ for the distribution $H.$

Other local policy effects can be considered by specifying $\theta (F)$ to
be something other than a feature of the distribution of a random variable $%
Y $. One example of such a $\theta (F)$ is a bound on average equivalent
variation from Hausman and Newey (2016). The Gateaux derivative formula in
equation (\ref{G deriv}) can be used to derive local policy effects of
structural changes on this and many other objects. In Section 3 we do so for 
$\theta (F)$ that depends on conditional location.

Specification and estimation of global policy effects using quantile
regressions was developed by Machado and Mata (2005), Albrecht, Bj{\"{o}}%
rklund, Vroman (2003), and Melly (2005). Estimators of global effects based
distribution regression were developed by Chernozhukov, Fernandez-Val, and
Melly (2013). Local policy effects are useful for evaluating small policies.
Also, Muhkin (2019) shows that global policy effects can be obtained from
integrating local effects, making local effects of interest even for
evaluation of global effects.

\subsection{Local Sensitivity and Local Hausman Tests}

Quantifying local sensitivity of an estimator to misspecification, or more
generally to a change in distribution of the data, is another important use
of the influence function. Equation (\ref{G deriv}) gives the Gateaux
derivative of the probability limit $\theta (F)$ in the direction $H-F_{0}.$
If the distribution $H$ allows for misspecification then $\int \psi (w)H(dw)$
measures sensitivity of $\theta (F)$ to local misspecification. More
generally if $H$ is a different distribution than that of the data then $%
\int \psi (w)H(dw)$ measures the sensitivity of $\theta (F)$ to a
distribution shift. Qualitative and quantitative sensitivity measures can be
constructed based on $\psi (w).$ A qualitative sensitivity characteristic is
boundedness of $\psi (w)$, which guarantees that $d\theta (F_{\tau })/d\tau $
is bounded over all possible $H.$ This is the classic robustness
characteristic of Hampel (1974) and Huber (1981) that is defined by
boundedness of the influence function.

Quantitative measures of estimator sensitivity can also be based on $\psi
(w).$ Conley, Hansen, and Rossi (2012) and Andrews, Gentzkow, and Shapiro
(2017) give measures of sensitivity of IV and GMM estimators, respectively,
to moment misspecification. The sensitivity measure for GMM is exactly $\int
\psi (w)H(dw)$ for the GMM influence function. To explain, suppose that
there is a vector function $g(w,\theta )$ of a data observation $w$ and
parameter vector $\theta $ satisfying a moment condition $E[g(W,\theta
_{0})]=0.$ A GMM estimator is obtained as $\hat{\theta}=\arg \min_{\theta }%
\hat{g}(\theta )^{\prime }\hat{\Psi}\hat{g}(\theta )$ where $\hat{g}(\theta
)=\sum_{i=1}^{n}g(W_{i},\theta )/n$ are sample moments and $\hat{\Psi}$ is a
positive semi-definite weighting matrix. It is well known that the influence
function for GMM under correct specification (i.e. $E[g(W,\theta _{0})]=0$)
is 
\begin{equation*}
\psi (w)=-(G^{\prime }\Psi G)^{-1}G^{\prime }\Psi g(w,\theta _{0}),\text{ }%
G=\left. \frac{\partial }{\partial \theta }E[g(W,\theta )]\right\vert
_{\theta =\theta _{0}},\text{ }\Psi =\text{plim}(\hat{\Psi}).\text{ }
\end{equation*}%
Therefore for GMM the local sensitivity will be%
\begin{equation*}
\frac{d\theta (F_{\tau })}{d\tau }=\int \psi (w)H(dw)=-(G^{\prime }\Psi
G)^{-1}G^{\prime }\Psi \int g(w,\theta _{0})H(dw).
\end{equation*}%
This is the local sensitivity formula given in Andrews, Gentzkow, and
Shapiro (2017). When the dimension of $g(w,\theta )$ is bigger than that of $%
\theta $ this formula imposes correct specification of the moments, i.e. $%
E[g(W,\theta _{0})]=0$. Imbens (1997) gives the influence function for GMM
allowing for misspecification and Muhkin (2019) describes its use for
sensitivity analysis.

Equation (\ref{G deriv}) gives the local sensitivity of any estimator to a
change of $F$ in the direction $H-F_{0}$. In Section 3 we derive local
sensitivity of a functional of conditional location and illustrate its use
in estimating sensitivity of average equivalent variation bounds to
endogeneity of gasoline prices.

Local sensitivity can be used to construct local Hausman specification tests
for any object of interest with an influence function. A first order
expansion gives%
\begin{equation}
\theta (H)-\theta (F_{0})=\theta (F_{1})-\theta (F_{0})\approx \left( \frac{%
d\theta (F_{\tau })}{d\tau }(\tau -0)\right) _{\tau =1}=\int \psi (w)H(dw).
\label{haus}
\end{equation}%
Thus we see that $\int \psi (w)H(dw)$ is a first order approximation to the
effect of changing the distribution $F$ on the probability limit $\theta (F)$
of the estimator $\hat{\theta}$ corresponding to Hausman's (1978) idea of
checking sensitivity of an estimator of an object of interest to model
assumptions. A estimator of $d\theta (F_{\tau })/d\tau $ can be formed from
an estimator of the influence function $\psi (w)$ and an alternative $H$ by
substituting the estimated influence function in equation (\ref{G deriv})
and integrating over $H$. Standard errors can be constructed using
asymptotic theory or the bootstrap and an asymptotic t-statistic formed in
the usual way. From equation (\ref{haus}) we see that such a t-statistic is
a local Hausman test of the effect of misspecification in the direction $H.$%
. This approach can give local Hausman specification tests for any estimator
with an influence function in any direction $H$. In Section 3 we illustrate
such tests by testing for a significant effect of endogeneity of price on
average equivalent variation for gasoline demand. It is beyond the scope of
this paper to develop the general asymptotic theory of such tests. We
discuss these tests here to illustrate the usefulness of the influence
function in empirical work.

The covariance between the influence functions of two different estimators
was suggested by Gentzkow and Shapiro (2015) and Andrews, Gentzkow, and
Shapiro (2017) as a measure of sensitivity of one estimator with respect to
another. Muhkin (2019) gives a geometric interpretation of this covariance
as a directional derivative of one functional with respect another. As
Muhkin (2019) shows, the covariance between two influence functions is the
Gateaux derivative of $\theta (F)$ with respect to a departure from $F_{0}$
in a direction $G$ that corresponds to a change in the other functional. In
this way the influence functions for two different estimators are useful for
constructing measures of sensitivity. For brevity we omit further specifics
but note that this is an active and important research topic that is
potentially useful for empirical work, where influence functions are key
ingredients.

\subsection{Orthogonal Moment Functions}

Another important use of influence functions is in the construction of
orthogonal moment functions for GMM with a nonparametric first step.
Orthogonal moment functions are those where the expected moment functions
have zero derivative with respect to the first step. GMM with orthogonal
moment functions does not suffer from the large model selection and
regularization biases of some estimators based on nonorthogonal moment
functions. Avoiding such biases can be particularly important for machine
learning first steps, as discussed in Chernozhukov et al. (2018) and shown
in Chernozhukov et al. (2020).

To describe orthogonal moment functions consider a vector of functions $%
g(w,\gamma ,\theta )$ where $\gamma $ is a (possibly) nonparametric first
step with true value $\gamma _{0}$, $\theta $ is the parameter vector of
interest, and the moment condition $E[g(W,\gamma _{0},\theta _{0})]=0$ is
satisfied. This moment condition can be thought of as an identifying moment
for $\theta _{0},$ with $\gamma _{0}$ obtained from a first step. In general
the first order effect of $\gamma $ on $E[g(W,\gamma ,\theta _{0})]$ may be
nonzero, leading to bias in a GMM estimator based on sample moments $\hat{g}%
(\theta )=\sum_{i=1}^{n}g(W_{i},\hat{\gamma},\theta )/n$, where $\hat{\gamma}
$ is a first step estimator of $\gamma _{0}$ that is plugged in. As shown in
Chernozhukov et al. (2020), orthogonal moment functions can be constructed
by adding to the identifying moments the influence function $\phi (w,\gamma
_{0},\alpha _{0},\theta )$ of $E[g(W,\gamma (F),\theta )]$, where $\alpha $
are additional unknown functions on which $\phi $ may depend and $\gamma (F)$
is the probability limit of the first step estimator $\hat{\gamma}\,$when $F$
is the true distribution of $W$. This $\phi (w,\gamma ,\alpha ,\theta )$ can
be calculated by Steps I and II applied to equation (\ref{G deriv}) for $%
E[g(W,\gamma (F),\theta )]$, i.e%
\begin{equation}
\frac{dE[g(W,\gamma (F_{\tau }),\theta )]}{d\tau }=\int \phi (w,\gamma
_{0},\alpha _{0},\theta )H(dw),\text{ }E[\phi (W,\gamma _{0},\alpha
_{0},\theta )]=0.  \label{adj term}
\end{equation}%
Orthogonal moment functions can then be constructed as%
\begin{equation*}
\psi (w,\gamma ,\alpha ,\theta )=g(w,\gamma ,\theta )+\phi (w,\gamma ,\alpha
,\theta ).
\end{equation*}

The influence function $\phi (w,\gamma _{0},\alpha _{0},\theta )$ of $%
E[g(W,\gamma (F),\theta )]$ is an "adjustment term," analyzed in Newey
(1994), that accounts for the presence of the first step $\hat{\gamma}$ in
the moment functions. We here refer this this adjustment term as the first
step influence function (FSIF). Adding the FSIF to the original, identifying
moment functions $g(w,\gamma ,\theta )$ makes orthogonal moments.
Calculating the FSIF from Steps I and II is simpler than obtaining $\phi $
from the functional equation in Newey (1994). This simplicity facilitates
the construction of orthogonal moment functions. We illustrate by
calculating the FSIF $\phi $ for solutions to exogenous orthogonality
conditions in Section 3 and endogenous orthogonality conditions in Section
4. In Chernozhukov et al. (2020) the FSIF for quantile orthogonality
conditions is used to obtain debiased machine learning estimators for
functionals of solutions to quantile conditions.

Local policy analysis, sensitivity measures, and constructing orthogonal
moment functions are three uses of the influence function that are of direct
interest for empirical research. The results of this paper are useful in
providing a simpler method of calculating the influence function that can
then be used to construct local policy effects of structural changes, local
sensitivity analysis and local Hausman tests for any estimator with an
influence function, and orthogonal moment functions that can be used in
debiased machine learning. In the next Section we illustrate by deriving the
influence function for conditional location effects, constructing
sensitivity measures for estimators of such effects, and applying them to
average equivalent variation bounds.

Another important use of the influence function is in asymptotic efficiency
comparisons, where it is convenient to bypass the stochastic expansion in
equation (\ref{inf}). Knowing the influence function is also useful for
showing that the asymptotic expansion in equation (\ref{inf}) is satisfied,
because the influence function implies the precise form of the remainder.
For brevity we omit further discussions of these uses of the influence
function.

\section{Exogenous Orthogonality Conditions}

Many interesting economic and causal effects depend on a function that
solves an orthogonality condition and depends only on exogenous instrumental
variables. Such functions include high dimensional or additive
specifications of orthogonality conditions for quantiles or expectiles.
Effects of interest include bounds on average equivalent variation and
average derivatives. In this Section we derive the influence function for
such effects using Step I\ and Step II. We quantify local policy effects and
local sensitivity for these effects. In addition we give an application to
sensitivity of bounds on average equivalent variation to endogeneity in
gasoline demand.

\subsection{Functions Satisfying Exogenous Orthogonality Conditions}

The unknown functions we consider depend on a vector of regressors $X$ that
may be infinite dimensional. We will denote a possible unknown function by $%
\gamma $ with $\gamma (x)$ being its realization at $X=x.$ We will impose
the restriction that $\gamma $ is in a set of functions $\Gamma $ that is
linear and closed in mean square, meaning that every $\gamma $ in $\Gamma $
has finite second moment and that if $\gamma _{k}\in \Gamma $ for each
positive integer $k$ and $E[\{\gamma _{k}(X)-\gamma
(X)\}^{2}]\longrightarrow 0$ then $\gamma \in \Gamma .$ We give examples of $%
\Gamma $ in the second paragraph to follow.

We specify $\gamma _{0}=\gamma (F_{0})$ to be the probability limit (plim)
of a nonparametric estimator $\hat{\gamma}$ when the distribution is $F_{0}.$
We suppose that $\gamma _{0}$ satisfies an orthogonality condition where a
residual $\rho (W,\gamma )$ with finite second moment is orthogonal in the
population to all $b\in \Gamma .$ That is we specify that $\gamma _{0}$
satisfies 
\begin{equation}
E[b(X)\rho (W,\gamma _{0})]=0\text{ for all }b\in \Gamma .  \label{exog orth}
\end{equation}%
This is like an instrumental variables orthogonality condition where the
function $\gamma $ depends only on the same variables $X$ that the
instrumental variables $b(X)$ depend on. This dependence of the functions $%
\gamma $ and instrumental variables $b$ on the same $X$ is the "exogenous"
referred to in the title of this Section. In the next Section we consider
orthogonality conditions where $\gamma $ may depend on different variables
than $X$, corresponding to instrumental variables settings where there is
endogeneity.

If $\Gamma $ is specified to be all functions of $\Gamma $ with finite
second moment then equation (\ref{exog orth}) will be a conditional moment
restriction $E[\rho (W,\gamma _{0})|X]=0.$ We also allow $\Gamma $ to be a
smaller set. For example, a set of functions of interest for high
dimensional estimation are those that are linear combinations of a sequence
of functions $(b_{1}(X),b_{2}(X),...)$ each having finite second moment. A
corresponding $\Gamma $ would be limits in mean square of linear
combinations $\sum_{j=1}^{\infty }\beta _{j}b_{j}(X)$ where $\beta _{j}\neq
0 $ for only a finite number of \thinspace integers $j.$ Another example is
a set of functions that are additive in distinct components of $X$. For $%
X=(X_{1},X_{2})$ this $\Gamma $ is the mean square closure of all functions $%
\gamma (X)=\gamma _{1}(X_{1})+\gamma _{2}(X_{2})$ that are additive in in $%
X_{1}$ and $X_{2}$ with finite second moment. The high dimensional,
additive, and unrestricted specifications of $\Gamma $ are each of interest.

A leading example of the residual function is $\rho (W,\gamma )=Y-\gamma (X)$
for an outcome variable $Y$ having finite second moment. In this example the
orthogonality condition of equation (\ref{exog orth}) specifies that $\gamma
_{0}$ is the least squares projection of $Y$ on the set of functions $\Gamma 
$, i.e. $\gamma _{0}=\arg \min_{\gamma \in \Gamma }E[\{Y-\gamma (X)\}^{2}]$.
In this example $\gamma _{0}$ is the conditional expectation if $\Gamma \,$%
is all functions of $X$ with finite second moment, or is the least squares
projection of $Y$ on the closure of linear combinations of $%
(b_{1}(X),b_{2}(X),...)$, or is the least squares projection on the closure
of additive functions. Newey (1994) gives the influence function for
functionals of such $\gamma _{0}.$

There are other important examples of the residual function.

\bigskip

\textsc{Quantile:} In this case there is an outcome variable $Y$ and the
residual function is 
\begin{equation*}
\rho (W,\gamma )=p-1(Y<\gamma (X)),
\end{equation*}%
where $0<p<1.$ This $\rho (W,\gamma )$ is the derivative with respect to $u$
of the "check function" $q_{p}(u)=\left\vert u\right\vert
\{p1(u>0)+(1-p)1(u<0)\}$ evaluated at $u=Y-\gamma (X);$ see Koenker and
Bassett (1978). By convexity of $q_{p}(Y-\gamma (X))$ in $\gamma $,%
\begin{equation*}
\gamma _{0}=\arg \min_{\gamma \in \Gamma }E[q_{p}(Y-\gamma (X))].
\end{equation*}%
Here $\gamma _{0}(X)$ will the $p^{th}$ conditional quantile of $Y$ when $%
\Gamma $ is unrestricted. For other specifications of $\Gamma $ the $\gamma
_{0}$ will be minimum of the expected check function over $\Gamma $.

\bigskip

\textsc{Expectile:} In this case the residual function is%
\begin{equation*}
\rho (W,\gamma )=[p+(1-2p)1(Y<\gamma (X))](Y-\gamma (X)).
\end{equation*}%
This $\rho (W,\gamma )$ is the derivative with respect to $u$ of the
asymmetric squared residual function $\bar{q}_{p}(u)=(u^{2}/2)%
\{p1(u>0)+(1-p)1(u<0)\}$ evaluated at $u=Y-\gamma (X),$ as in Newey and
Powell (1987). By convexity of $\bar{q}_{p}(Y-\gamma (X))$ in $\gamma $%
\begin{equation*}
\gamma _{0}=\arg \min_{\gamma \in \Gamma }E[\bar{q}_{p}(Y-\gamma (X))].
\end{equation*}%
Here $\gamma _{0}(X)$ will the $p^{th}$ conditional expectile of $Y$ given $%
X $ when $\Gamma $ is unrestricted. For other specifications of $\Gamma $
the $\gamma _{0}$ will be minimum of the asymmetric squared residual
function over $\Gamma $.

\bigskip

\textsc{Binary choice:} In this case there is a binary outcome variable $%
Y\in \{0,1\},$ a known CDF $\Lambda (a)\,$\ with derivative (pdf) $\Lambda
_{a}(a)$, and the residual is%
\begin{equation*}
\rho (W,\gamma )=\frac{\Lambda _{a}(\gamma (X))}{\Lambda (\gamma
(X))[1-\Lambda (\gamma (X))]}\{Y-\Lambda (\gamma (X))\}.
\end{equation*}%
This $\rho (W,\gamma )$ is $\partial Q(W,a)/\partial a$ at $a=\gamma (X)$
for the negative of the binary pseudo-likelihood 
\begin{equation*}
Q(W,a)=-Y\ln \Lambda (a)-(1-Y)\ln [1-\Lambda (a)].
\end{equation*}%
When $\ln (\Lambda _{a}(a))$ is concave this $Q(W,a)$ will be convex in $a$,
see Pratt (1981). For example the logit CDF $\Lambda (a)=e^{a}/[1+e^{a}]$
has this property with $\Lambda _{a}(a)/\{\Lambda (a)[1-\Lambda (a)]\}=1.$
The $\gamma _{0}$ will satisfy%
\begin{equation*}
\gamma _{0}=\arg \min_{\gamma \in \Gamma }E[Q(W,\gamma (X))].
\end{equation*}%
Here $\gamma _{0}(X)$ will be $\Lambda ^{-1}(\Pr (Y=1|X))$ when $\Gamma $ is
unrestricted. For other specifications of $\Gamma $ the $\gamma _{0}$ will
minimize the expected value of the negative log-likelihood $E[-Y\ln (\Lambda
(\gamma (X)))-(1-Y)\ln \{1-\Lambda (\gamma (X))\}]$ over $\Gamma $.

\bigskip

These cases of the residual function have the common feature that $\rho
(W,\gamma )=\left. dQ(W,a)/da\right\vert _{a=\gamma (X)}$ where $Q(W,a)$ is
a convex function. In all such cases equation (\ref{exog orth}) will be the
necessary and sufficient first order condition for%
\begin{equation*}
\gamma _{0}=\arg \min_{\gamma \in \Gamma }E[Q(W,\gamma )],
\end{equation*}%
when the argmin exists and some regularity conditions are satisfied. We
focus on the orthogonality condition because it is potentially more general.

\subsection{The Influence Function}

We derive the influence function of objects of the form%
\begin{equation}
\theta (F)=E_{F}[m(W,\gamma (F))],\text{ }E_{F}[b(X)\rho (W,\gamma (F))]=0%
\text{ for all }b\in \Gamma .  \label{func}
\end{equation}%
Here the object of interest is the expectation of the function $m(W,\gamma )$
at $\gamma _{0}$. One example of this $\theta (F)$ is a bound on average
equivalent variation discussed in Section 3.4 to follow. Other examples will
be discussed later in this Section.

The influence function of $\theta (F)$ will be the sum of two terms. To
explain let $F_{\tau }=(1-\tau )F_{0}+\tau H$ $=F_{0}+\tau (H-F_{0}),$ $%
0<\tau <1,$ denote a convex combination of the true CDF $F_{0}$ with another
CDF $H$ as discussed in Section 2 and let $\gamma _{\tau }=\gamma (F_{\tau })
$ and $E_{\tau }[\cdot ]=E_{F_{\tau }}[\cdot ]$. By the chain rule of
calculus,%
\begin{eqnarray*}
\frac{\partial }{\partial \tau }\theta (F_{\tau }) &=&\frac{\partial }{%
\partial \tau }E_{\tau }[m(W,\gamma _{0})]+\frac{\partial }{\partial \tau }%
E[m(W,\gamma _{\tau })] \\
&=&\int m(w,\gamma _{0})\{H-F_{0}\}(dw)+\frac{\partial }{\partial \tau }%
E[m(W,\gamma _{\tau })] \\
&=&\int [m(w,\gamma _{0})-\theta _{0}]H(dw)+\frac{\partial }{\partial \tau }%
E[m(W,\gamma _{\tau })].
\end{eqnarray*}%
We see in this equation that influence function of $\theta (F)$ will be the
sum of $m(w,\gamma )-\theta $ and a term $\phi (w,\gamma ,\alpha )$
satisfying%
\begin{equation}
\frac{\partial }{\partial \tau }E[m(W,\gamma _{\tau })]=\int \phi (w,\gamma
_{0},\alpha _{0})H(dw),  \label{npinf}
\end{equation}%
with 
\begin{equation*}
\frac{\partial }{\partial \tau }\theta (F_{\tau })=\int \psi (w,\gamma
_{0},\alpha _{0},\theta _{0})H(dw),\text{ }\psi (w,\gamma ,\alpha ,\theta
)=m(W,\gamma )-\theta +\phi (w,\gamma ,\alpha )
\end{equation*}%
The first term $m(w,\gamma )-\theta $ accounts for the unknown distribution $%
F$ that averages over $W$ in $m(W,\gamma _{0})-\theta _{0}.$ The second term 
$\phi (w,\gamma ,\alpha )$ accounts for estimation of the unknown $\gamma
_{0}$ satisfying the orthogonality condition of equation (\ref{exog orth}).
This $\phi (w,\gamma ,\alpha )$ is the FSIF, referred to as the "adjustment
term" in Newey (1994), that accounts for a nonparametric estimator of $%
\gamma _{0}$ satisfying equation (\ref{exog orth}). We focus here on the
derivation of $\phi (w,\gamma ,\alpha ).$

To derive $\phi (w,\gamma ,\alpha )$ we assume that $\gamma _{\tau }=\gamma
(F_{\tau })$ satisfies the orthogonality condition in equation (\ref{func})
for each $\tau $ so that for all $b\in \Gamma $%
\begin{equation}
E_{\tau }[b(X)\rho (W,\gamma _{\tau })]\equiv 0,  \label{orth cond par}
\end{equation}%
identically in $\tau $. We are implicitly assuming here that $\mathcal{%
\Gamma }$ does not depend on $\tau $ which will hold for the $F_{\tau }$ of
Appendix A. Differentiating this identity with respect to $\tau $ and
applying the chain rule of calculus, so that the derivative is the sum of
derivatives with respect to $\tau $ in $E_{\tau }[b(X)\rho (W,\gamma _{0})]$
and $E[b(X)\rho (W,\gamma _{\tau })],$ gives%
\begin{eqnarray}
0 &=&\frac{\partial }{\partial \tau }E_{\tau }[b(X)\rho (W,\gamma _{0})]+%
\frac{\partial }{\partial \tau }E[b(X)\rho (W,\gamma _{\tau })]
\label{orth cond der} \\
&=&\int b(x)\rho (w,\gamma _{0})H(dw)+\frac{\partial }{\partial \tau }%
E[b(X)\rho (W,\gamma _{\tau })],\text{ for all }b\in \Gamma .  \notag
\end{eqnarray}%
Solving gives%
\begin{equation*}
-\frac{\partial }{\partial \tau }E[b(X)\rho (W,\gamma _{\tau })]=\int
b(x)\rho (w,\gamma _{0})H(dw),\text{for all }b\in \Gamma .
\end{equation*}%
The object being integrated on the right provides a candidate for FSIF $\phi
(w,\gamma ,\alpha )$. This equation will give us equation (\ref{npinf}) if
there is $\alpha _{0}\in \Gamma $ with 
\begin{equation}
\frac{\partial }{\partial \tau }E[m(W,\gamma _{\tau })]=-\frac{\partial }{%
\partial \tau }E[\alpha _{0}(X)\rho (W,\gamma _{\tau })].
\label{alpha ident}
\end{equation}%
Such an $\alpha _{0}(X)$ will exist under the following two conditions.

\bigskip

\textsc{Assumption 1: }\textit{There exists }$v_{m}(X)$\textit{\ such that }$%
\partial E[m(W,\gamma _{\tau })]/\partial \tau =\partial E[v_{m}(X)\gamma
_{\tau }(X)]/\partial \tau $ and $E[v_{m}(X)^{2}]<\infty .$

\bigskip

Generally it will follow from the chain rule, iterated expectations, and $%
E[m(W,\gamma +a)|X]$ differentiable in a scalar $a$ that%
\begin{equation*}
v_{m}(X)=\left. \frac{\partial }{\partial a}E[m(W,\gamma
_{0}+a)|X]\right\vert _{a=0}.
\end{equation*}%
Assumption 1 is like equation (4.4) of Newey (1994) in requiring that $%
\partial E[m(W,\gamma _{\tau })]/\partial \tau $ can be represented as the
derivative of an expected product of a function $v_{m}(X)$ with $\gamma
_{\tau }(X)$ where $v_{m}(X)$ has finite second moment. One example is $%
m(W,\gamma )=v_{m}(X)\gamma (X)$ where $m(W,\gamma )$ is simply the product
of some function $v_{m}(X)$ with $\gamma (X)$ and the $v_{m}(X)$ of
Assumption 1 is the same as $v_{m}(X)$ here. Assumption 1 is also satisfied
for other important effects as further discussed below. In general this
condition with $E[v_{m}(X)^{2}]<\infty $ can be shown to be a necessary
condition for $\theta (F)$ to have a finite semiparametric variance bound.

\bigskip

\textsc{Assumption 2: }There is $v_{\rho }(X)<0$ that is bounded and bounded
away from zero such that $\partial E[b(X)\rho (W,\gamma _{\tau })]/\partial
\tau =\partial E[b(X)v_{\rho }(X)\gamma _{\tau }(X)]/\partial \tau $ for
every $b\in \Gamma $.

\bigskip

Generally it will follow from the chain rule, iterated expectations, and $%
E[\rho (W,\gamma _{0}+a)|X]$ differentiable in a scalar $a$ that%
\begin{equation*}
v_{\rho }(X)=\left. \frac{\partial }{\partial a}E[\rho (W,\gamma
_{0}+a)|X]\right\vert _{a=0}
\end{equation*}%
In this way Assumption 2 allows for $\rho (W,\gamma )$ to not be continuous
as long as $E[\rho (W,a)|X]$ is differentiable in $a$. Here $v_{\rho }(X)<0$
is a sign normalization while $v_{\rho }(X)$ being bounded and bounded away
from zero is important for the results. For example $v_{\rho }(X)=-1$ for $%
\rho (W,\gamma )=Y-\gamma (X).$

Under Assumptions 1 and 2 equation (\ref{alpha ident}) becomes%
\begin{equation*}
\frac{\partial }{\partial \tau }E[v_{m}(X)\gamma _{\tau }(X)]=-\frac{%
\partial }{\partial \tau }E[\alpha _{0}(X)v_{\rho }(X)\gamma _{\tau }(X)].
\end{equation*}%
This equality will be satisfied if $E[v_{m}(X)\gamma _{\tau }(X)]=E[\alpha
_{0}(X)v_{\rho }(X)\gamma _{\tau }(X)]$ for all $\tau .$ Since $\gamma
_{\tau }\in \Gamma $ this condition will be satisfied if for all $\gamma \in
\Gamma ,$%
\begin{equation*}
E[v_{m}(X)\gamma (X)]=-E[\alpha _{0}(X)v_{\rho }(X)\gamma (X)].
\end{equation*}%
Adding $E[\alpha _{0}(X)v_{\rho }(X)\gamma (X)]$ to both sides gives%
\begin{eqnarray*}
0 &=&E[v_{m}(X)\gamma (X)]+E[\alpha _{0}(X)v_{\rho }(X)\gamma (X)] \\
&=&E[\{-v_{\rho }(X)\}\{\frac{-v_{m}(X)}{v_{\rho }(X)}-\alpha
_{0}(X)\}\gamma (X)]\text{ for all }\gamma \in \Gamma ,
\end{eqnarray*}%
where the second equality follows by multiplying and dividing by $-v_{\rho
}(X)$ in $E[v_{m}(X)\gamma (X)]$. This is the orthogonality condition that
is necessary and sufficient for $\alpha _{0}(X)$ to be the weighted least
squares projection of $-v_{m}(X)/v_{\rho }(X)$ on $\Gamma $ for weight $%
-v_{\rho }(X)$.

\bigskip

\textsc{Proposition 1}: \textit{If Assumptions 1 and 2 are satisfied then }%
\begin{equation}
\phi (w,\gamma ,\alpha )=\alpha (x)\rho (w,\gamma ),\text{ }\alpha
_{0}(x)=\arg \min_{\alpha \in \Gamma }E[\{-v_{\rho }(X)\}\{-v_{m}(X)/v_{\rho
}(X)-\alpha (X)\}^{2}].  \notag
\end{equation}

\bigskip

Proposition 1 generalizes Proposition 4 of Newey (1994) where $\phi
(w,\gamma ,\alpha )$ was given for least squares projections where $\rho
(W,\gamma )=Y-\gamma (X).$ Here we give the FSIF $\phi (w,\gamma ,\alpha )$
for any plim $\gamma (F)$ of a first step $\hat{\gamma}$ satisfying the the
exogenous orthogonality condition of equation (\ref{func}) where Assumptions
1 and 2 are also satisfied. We have obtained Proposition 1 by
differentiation the orthogonality condition (\ref{orth cond par}) with
respect to $\tau $ and choosing the instrumental variable $b(X)$ in that
condition so that equation (\ref{alpha ident}) is satisfied. This derivation
of Proposition 1 illustrates how the FSIF can be {obtained directly from the
moment conditions defining the first step estimator without solving an
integral equation} or {using asymptotic arguments. }

First steps that solve orthogonality conditions for quantiles, expectiles,
and binary choice provide useful examples.

\bigskip

\textsc{Example 1:} \textit{Quantile Functional; }For $\rho (W,\gamma
)=p-1(Y<\gamma (X))$%
\begin{equation*}
-v_{\rho }(X)=\frac{\partial \Pr (Y<\gamma _{0}(X)+a|X)]}{\partial a}%
=f_{Y|X}(\gamma _{0}(X)|X)
\end{equation*}%
where $f_{Y|X}(y|X)$ is the pdf of $Y$ conditional on $X.$ The FSIF is%
\begin{equation*}
\phi (w,\gamma ,\alpha )=\alpha (X)[p-1(y<\gamma (x))],
\end{equation*}%
where $\alpha _{0}$ is given in Proposition 1. The formula for $\alpha _{0}$
depends on the functional $m(W,\gamma )$ through the derivative term $%
v_{m}(W)$ and is given by%
\begin{equation*}
\alpha _{0}(x)=\arg \min_{\alpha \in \Gamma }E[f_{Y|X}(\gamma
_{0}(X)|X)\{v_{m}(X)/f_{Y|X}(\gamma _{0}(X)|X)-\alpha (X)\}^{2}].
\end{equation*}%
For instance consider a weighted average derivative functional where $%
m(W,\gamma )=w(x)\partial \gamma (x)/\partial x_{1}.$ Integration by parts
gives%
\begin{eqnarray*}
E[m(W,\gamma )] &=&\int w(x)\frac{\partial \gamma (x)}{\partial x}%
f_{0}(x)dx=-\int \frac{\partial \{w(x)f_{0}(x)\}}{\partial x_{1}}\gamma
(x)dx=E[v_{m}(X)\gamma (X)], \\
v_{m}(X) &=&-\frac{1}{f_{0}(X)}\frac{\partial \{w(X)f_{0}(X)\}}{\partial
x_{1}}.
\end{eqnarray*}%
When $\Gamma $ is unrestricted Proposition 1 gives $\alpha
_{0}(X)=v_{m}(X)/f_{Y|X}(\gamma _{0}(X)|X)$ and the FSIF coincides with that
of Chauduri, Doksum, and Tsybakov (1997). Ackerberg et al. (2014) also gave
an expression for the FSIF for quantile functionals other than the weighted
average derivative with $v_{m}(X)$ replaced by a functional derivative of $%
E[m(W,\gamma )]$. When $\Gamma $ is restricted then $\alpha _{0}(X)$ being
the weighted projection of $v_{m}(X)/f_{Y|X}(\gamma _{0}(X)|X)$ on $\Gamma $
with weight $f_{Y|X}(\gamma _{0}(X)|X).$ Proposition 1 generalizes the
previous results to allow restrictions on $\gamma .$

\bigskip

\textsc{Example 2:} \textit{Expectile Functional; }For a conditional
expectile $\rho (W,\gamma )=[p1(Y>\gamma (X))+(1-p)1(Y<\gamma (X))][Y-\gamma
(X)],$ so that%
\begin{equation*}
-v_{\rho }(X)=p\Pr (Y>\gamma _{0}(X)|X)+(1-p)\Pr (Y<\gamma _{0}(X)|X),
\end{equation*}%
which is bounded and bounded away from zero. The FSIF is%
\begin{equation*}
\phi (w,\gamma ,\alpha )=-\alpha (X)[p1(Y>\gamma (X))+(1-p)1(Y<\gamma
(X))][Y-\gamma (X)],
\end{equation*}%
where $\alpha _{0}(X)$ is given in Proposition 1. The formula for $\alpha
_{0}$ depends on the functional $m(W,\gamma )$ through the derivative term $%
v_{m}(W)$ and is given. When $\Gamma $ is unrestricted and $m(W,\gamma
)=w(x)\partial \gamma (x)/\partial x_{1}$ then $v_{m}(X)$ will be as in
Example 1 and $\alpha _{0}(X)=-v_{m}(X)/v_{\rho }(X).$ We are not aware of
previous results on the FSIF for functions that minimize the expectile
objective function.

\bigskip

Examples 1 and 2 illustrate how the term $v_{m}(X)$ is determined by the
functional of interest while $v_{\rho }(X)$ is determined by the residual $%
\rho (W,\gamma ).$ Proposition 1 shows how these aspects are combined to
determine the $\alpha _{0}(X)\in \mathcal{B}$ that multiplies the residual $%
\rho (W,\gamma )$ to form the FSIF. From equation (\ref{alpha ident}) we see
that this $\alpha _{0}(X)$ is precisely the function that makes the effect
of $\gamma _{\tau }$ on $E[m(W,\gamma _{\tau })]$ equal to the effect of $%
\gamma _{\tau }$ on $-E[\alpha _{0}(X)\rho (W,\gamma _{\tau })].$
Proposition 1 shows that this $\alpha _{0}(X)$ is a projection of $%
-v_{m}(X)/v_{\rho }(X)$ on $\Gamma $ weighted by $-v_{\rho }(X)$

The explicit formula in Proposition 1 is useful for quantifying local policy
effects and local sensitivity of semiparametric estimators, as we will
illustrate in the remainder of this Section. Proposition 1 also illustrates
how the influence function can be obtained with calculus, under natural
differentiability conditions like Assumptions 1 and 2. The key steps in
deriving Proposition 1 are to use the first order condition for $\gamma (F)$
to derive candidates for the influence function and to show that equation (%
\ref{npinf}) is satisfied for one of those candidates.

\subsection{Generalizing the Omitted Variable Bias Formula}

The influence function for exogenous orthogonality conditions can be used to
quantify local sensitivity to distributional changes of any object with an
influence function. We consider structural changes where the distribution of 
$X$ remains the same but the distribution of the outcome variable $Y$ given $%
X$ is different. A leading example, as we will see, is the omitted variable
problem. We focus on the case where $m(w,\gamma )$ depends only on $x$,
which covers many examples of interest and leads to simple, intuitive
formulas. We consider $H$ where the marginal distribution of $X$ is the same
as for $F_{0}$ but $\rho (W,\gamma _{0})$ may not be orthogonal to $\Gamma .$
Because $E_{H}[m(W,\gamma _{0})]=E[m(X,\gamma _{0})]=$ $\theta _{0}$ the
local sensitivity to such $H$ is given by the following result:

\bigskip

\textsc{Proposition 2:} \textit{If Assumptions 1 and 2 are satisfied, }$%
m(W,\gamma _{0})$\textit{\ depends only on }$X,$\textit{\ and }$H$\textit{\
has the same marginal distribution of }$X$\textit{\ as }$F_{0}$\textit{\
then } 
\begin{equation}
\frac{d\theta (F_{\tau })}{d\tau }=E_{H}[\alpha _{0}(X)\rho (W,\gamma _{0})].
\label{ce sens}
\end{equation}

\bigskip

Here we see that the local sensitivity is the expected product of $\alpha
_{0}(X)$ with the conditional mean of the residual $\rho (W,\gamma _{0})$
under the alternative distribution $H.$ This local sensitivity formula
generalizes the classic omitted variable bias formula to the local bias of
any object that depends on the solution to an exogenous orthogonality
condition, as we now demonstrate.

\bigskip

\textsc{Example 3: }\textit{Omitted Variable Bias Formula;} Here we show
that the classic omitted variable bias formula is a special case of
Proposition 2. Consider the conditional mean $\gamma _{0}(X)=E[Y|X]$ where $%
X $ has finite support and let $D$ be the indicator function of one of the
possible discrete outcomes of $X$. Then there is $Z$, $\theta _{0},$ and $%
\gamma _{0}$ such that%
\begin{equation*}
E[Y|D,Z]=\gamma _{0}(X)=D\theta _{0}+Z^{\prime }\gamma _{0}.
\end{equation*}%
Take the object of interest to be $\theta _{0}$. Let $\tilde{D}=D-E[D|Z]$ be
the residual from the population least squares regression of $D$ on $Z$.
Then the coefficient $\theta _{0}$ is a functional of $\gamma _{0}(X)$ given
by 
\begin{equation*}
\theta _{0}=E[\alpha _{0}(X)\gamma _{0}(X)],\text{ }\alpha _{0}(X)=\frac{%
\tilde{D}}{E[\tilde{D}^{2}]}.
\end{equation*}%
Let $\varepsilon :=$ $Y-\gamma _{0}(X)=\rho (W,\gamma _{0}).$ The
sensitivity is then 
\begin{equation*}
\frac{d\theta (F_{\tau })}{d\tau }=E[\alpha _{0}(X)E_{H}[Y-\gamma
_{0}(X)|X]]=\frac{E[\tilde{D}E_{H}[\varepsilon |X]]}{E[\tilde{D}^{2}]}.
\end{equation*}%
If there is an omitted variable $\tilde{Z}$ under $H$ so that the
distribution $\varepsilon $ is the same as $Y-\gamma _{0}(X)-\tilde{Z}$ then 
\begin{equation*}
\frac{d\theta (F_{\tau })}{d\tau }=E[\alpha _{0}(X)E_{H}[\varepsilon |X]]=%
\frac{E[\tilde{D}E_{H}[\tilde{Z}|X]]}{E[\tilde{D}^{2}]}.
\end{equation*}%
This formula is the classic omitted variables bias formula.

\bigskip

Example 3 shows that Proposition 2 generalizes the omitted variables bias
formula for one coefficient of a linear regression to any object that
depends on a solution to an exogenous orthogonality condition. We will
illustrate another use of the generalization by estimating the local
sensitivity of a bound on average equivalent variation to endogeneity of the
price in a gasoline demand application.

An estimator of the local sensitivity can be obtained from an estimator $%
\hat{\alpha}(x)$ of the term $\alpha _{0}(x)$ in the influence function and
from a specification $\hat{H}$ of the joint distribution of $X$ and $\rho
(W,\gamma _{0})$ under misspecification as 
\begin{equation*}
\widehat{\frac{d\theta (F_{\tau })}{d\tau }}=\int \left[ \hat{\alpha}(x)\rho
(w,\hat{\gamma})\right] \hat{H}(dw).
\end{equation*}%
Construction of a local Hausman test based on this object would require an
estimator of the asymptotic variance of the sensitivity $\widehat{d\theta
(F_{\tau })/d\tau }.$ It is beyond the scope of this paper to derive the
asymptotic variance of the sensitivity and construct a consistent estimator
of that asymptotic variance, although a bootstrap variance estimator could
be used and should prove valid. We will illustrate in the gasoline demand
example how this could be done.

An important part of $\widehat{d\theta (F_{\tau })/d\tau }$ is an estimator $%
\hat{\alpha}(x)$ of $\alpha _{0}(x)$ that appears in the FSIF of Proposition
1. Such an $\hat{\alpha}(x)$ can be constructed as in Chernozhukov et al.
(2020). Consider a dictionary of functions $b(x)=(b_{1}(x),...,b_{p}(x))^{%
\prime }$ with $b_{j}\in \Gamma $ for each $j.$ As discussed following
Assumption 1 differentiablity of $E[m(W,\gamma _{0}+a)|X]$ in the constant $a
$ will lead to%
\begin{eqnarray}
\frac{\partial }{\partial \tau }E[m(W,\gamma +\tau b_{j})] &=&E[\frac{%
\partial }{\partial \tau }E[m(W,\gamma _{0}+\tau
b_{j})|X]]=E[v_{m}(W)b_{j}(X)]  \label{ahat momm} \\
&=&E[-v_{\rho }(X)\{\frac{v_{m}(X)}{-v_{\rho }(X)}\}b_{j}(X)]=E[\{-v_{\rho
}(X)\}\alpha _{0}(X)b_{j}(X)]  \notag \\
&=&-E[\frac{\partial }{\partial \tau }E[\rho (W,\gamma _{0}+\tau
b_{j})|X]\alpha _{0}(X)]  \notag \\
&=&-E[\rho _{\gamma }(W,\gamma _{0})\alpha _{0}(X)b_{j}(X)],\text{ }%
(j=1,...,p),  \notag
\end{eqnarray}%
where the third equality is obtained by multiplying and dividing by $%
-v_{\rho }(X),$ the fourth by $\alpha _{0}(X)$ being as given in Proposition
1, the fifth by the discussion following Assumption 2, and the last equality
by differentiability of $\rho (W,\gamma +a)$ in a constant $a$ with $\rho
_{\gamma }(W,\gamma _{0})$ being the derivative. These are moment conditions
that can be used to estimate $\alpha _{0}(X)$ as a linear combination of the
dictionary functions. The idea is to replace expectations with sample
averages, $\gamma _{0}$ with an estimator $\hat{\gamma},$ $\alpha _{0}(X)$
with a linear combination $\pi ^{\prime }b(X),$ and then solve for an
estimator of $\pi .$ Let 
\begin{equation*}
\hat{M}=(\hat{M}_{1},...,\hat{M}_{p})^{\prime }\text{, }\hat{M}_{j}=\frac{%
\partial }{\partial \tau }\frac{1}{n}\sum_{i=1}^{n}m(W_{i},\hat{\gamma}+\tau
b_{j}),\text{ }\hat{G}=\frac{1}{n}\sum_{i=1}^{n}\rho _{\gamma }(W_{i},\hat{%
\gamma})b(X_{i})b(X_{i})^{\prime }.\text{ }
\end{equation*}%
Then a version of equation (\ref{ahat momm}) that replaces expectations with
sample moments, $\gamma _{0}$ by $\hat{\gamma}$, and has $\pi ^{\prime }b(X)$
in place of $\alpha _{0}(X)$ is $\hat{M}=-\hat{G}\pi .$ Solving for $\pi $
gives%
\begin{equation}
\hat{\alpha}(x)=\hat{\pi}^{\prime }b(x),\text{ }\hat{\pi}=-\hat{G}^{-1}\hat{M%
}.  \label{alphahat}
\end{equation}%
For quantile orthogonality conditions where $\rho (W,\gamma )$ is not
continuous one can use kernel weighting to construct $\hat{G}$ as in Example
2 of Chernozhukov et al. (2020).

For regression where $\rho (W,\gamma )=Y-\gamma (X)$ this $\hat{\alpha}(x)$
is the same as in Equation (6.2) from Newey (1994). For other choices of $%
\rho (W,\gamma )$ this $\hat{\alpha}(x)$ could be derived from series
expansions given in Ai and Chen (2007), Ackerberg, Chen, and Hahn (2012),
and Ackerberg et al. (2014) for conditional moment restrictions and Chen and
Liao\ (2015) more generally. Such interesting estimators of the FSIF would
be particularly useful when its form is not known. Here we rely on the
explicit moment condition for $\alpha _{0}(X)$ in equation (\ref{ahat momm})
that is a special case of the Chernozhukov et al. (2020).

\subsection{Sensitivity of Average Equivalent Variation for Gasoline Demand}

One object that depends on a conditional expectation is the Hausman and
Newey (2016) bound on average equivalent variation (AEV) for heterogenous
demand. This bound allows for completely general heterogeneity where the
demand function for each person can be unique to that person. The bound does
depend on preferences being independent of observed price and income, a
strong exogeneity restriction. Here we test the effect of dropping that
exogeneity restriction on AEV using the local sensitivity results we have
obtained.

An important motivation for this test is the difficulty of allowing for
endogeneity with general heterogeneity. Endogeneity can be allowed for using
control functions, as in Hausman and Newey (2016), but existence of control
functions imposes strong restrictions as in Blundell and Matzkin (2014).
Blundell, Horowitz, and Parey (2017) allow for endogeneity where there is an
instrument for price but restrict heterogeneity to be scalar where bounds on
AEV are not known. Here we take a different approach to allowing for
endogeneity, where we test for sensitivity to bounds on AEV to endogeneity.

To describe and carry out this test we first describe the AEV bound and
apply Proposition 1 to derive its influence function.

\bigskip

\textsc{Example 4: }\textit{Average Equivalent Variation Bound;} Here $Y$ is
the share of income spent on a commodity and $X=(P_{1},Z),$ where $P_{1}$ is
the price of the commodity and $Z$ includes income $Z_{1}$, prices of other
goods, and other observable variables affecting utility. Let $\check{p}_{1}<%
\bar{p}_{1}$ be lower and upper prices over which the price of the commodity
can change, $\kappa $ a bound on the income effect, and $\omega (z)$ some
weight function. The object of interest is%
\begin{equation}
\theta _{0}=E\left[ \omega (Z)\int_{\check{p}_{1}}^{\bar{p}_{1}}\left( \frac{%
Z_{1}}{u}\right) \gamma _{0}(u,Z)\exp (-\kappa \lbrack u-\check{p}_{1}])du%
\right] ,  \label{AEV}
\end{equation}%
where $u$ is a variable of integration. If individual heterogeneity in
consumer preferences is independent of $X$ and $\kappa $ is a lower (upper)
bound on the derivative of consumption with respect to income across all
individuals, then $\theta _{0}$ is an upper (lower) bound on the weighted
average over consumers and over the distribution of $Z$ of equivalent
variation for a change in the price of the first good from $\check{p}_{1}$
to $\bar{p}_{1}$.

This object is a special case of that considered in Proposition 1 where $%
v(u)=u^{2}/2,$ $\gamma _{0}(X)=E[Y|X]$, and $m(w,\gamma )$ depends only on $%
x $ and is given by%
\begin{equation*}
m(x,\gamma )=\omega (z)\int_{\check{p}_{1}}^{\bar{p}_{1}}(z_{1}/u)\gamma
(u,z)\exp (-\kappa \lbrack u-\check{p}_{1}])du.
\end{equation*}%
From the form of $E[m(X,\gamma )]$ and multiplying and dividing by the
conditional pdf $f(p_{1}|z)$ we find%
\begin{equation*}
\alpha _{0}(x)=f(p_{1}|z)^{-1}\omega (z)1(\check{p}_{1}<p_{1}<\bar{p}%
_{1})(z_{1}/p_{1})\exp (-\kappa \lbrack p_{1}-\check{p}_{1}]).
\end{equation*}%
where $f(p_{1}|z)$ is the conditional pdf of $P_{1}$ given $Z.$

\bigskip

We apply Example 4 to test sensitivity of a bound on AEV to endogeneity of
price using gasoline demand data in Hausman and Newey (2016, 2017) and
Blundell, Horowitz, and Parey (2017). We use the estimator $\hat{\alpha}(x)$
given in equation (\ref{alphahat}) for several choices of basis functions.
For an estimate of $\varepsilon =Y-\gamma (X)$ that allows for endogeneity
we use a linear instrumental variable estimator where the share equation has
a constant, ln(price), and ln(income) with the Blundell, Horowitz, and Parey
(2017) price instrument that is the distance from the Gulf of Mexico. We
take $\hat{\varepsilon}_{i},$ $(i=1,...,n)$ to be the residuals from the
linear instrumental variables estimation and the sensitivity estimator to be%
\begin{equation*}
\widehat{\frac{d\theta (F_{\tau })}{d\tau }}=\frac{1}{n}\sum_{i=1}^{n}\hat{%
\alpha}(X_{i})\hat{\varepsilon}_{i}.
\end{equation*}%
This sensitivity estimate will depart from zero when $\hat{\alpha}(X_{i})$,
which depends on the price variable, is correlated with the instrumental
variables residuals $\hat{\varepsilon}_{i}$. In this application we use the
delta method and standard calculations to obtain a standard error for the
sensitivity estimator.

We use gasoline demand data from the 2001 U.S. National Household
Transportation Survey (NHTS). This survey is conducted every 5-8 years by
the Federal Highway Administration. The survey is designed to be a
nationally representative cross section which captures 24-hour travel
behavior of randomly-selected households. Data collected includes detailed
trip data and household characteristics such as income, age, and number of
drivers. We restrict our estimation sample to households with either one or
two gasoline-powered cars, vans, SUVs and pickup trucks. We exclude Alaska
and Hawaii. We use daily gasoline consumption, monthly state gasoline
prices, and annual household income. The data we use consists of 8,908
observations. Note that the mean price of gasoline was \$1.33 per gallon
with the mean number of drivers in a household equal to 2.04.

We specify the weight function in the measure of AEV to be $\omega (Z)=1$
and consider a price change from the mean of price in the data to a price
that is 10 percent higher. We set $\kappa =0$ so that the sensitivity will
be for a lower bound on AEV when gasoline is a normal good (the income
effect is positive) for all consumers. For the basis function $b(x)$ used to
estimate $\hat{\alpha}(x)$ we consider bivariate linear, quadratic, and
cubic function in ln(price) and ln(income). Because their presence had
little effect on AEV estimates in Hausman and Newey (2016, 2017) we do not
use covariates here. We do use simulation to estimate the integral that
appears in $m(x,\gamma )$ in the bound. For $u_{i}$ uniformly distributed on 
$[\check{p}_{1},\bar{p}_{1}]$ the $\hat{\alpha}(x)$ is given by%
\begin{equation*}
\hat{\alpha}(x)=\hat{\pi}^{\prime }b(x),\text{ }\hat{\pi}=[%
\sum_{i=1}^{n}b(X_{i})b(X_{i})^{\prime }]^{-1}(\bar{p}_{1}-\check{p}%
_{1})\sum_{i=1}^{n}\left( \frac{Z_{1i}}{u_{i}}\right) b(u_{i},Z_{i}),
\end{equation*}%
where $x=(p_{1},z^{\prime })^{\prime }$ and $z_{1}$ is income.

Table 1 reports the sensitivity estimates and their standard errors for
linear, quadratic, and cubic specifications of $b(x).$

\begin{center}
\begin{tabular}{|c|c|c|}
\multicolumn{3}{c}{Table 1: AEV Sensitivity to Endogeneity} \\ \hline\hline
& Sensitivity & AEV Bound \\ \hline\hline
Linear & $1.44$ & $25.08$ \\ \cline{2-3}
& $(.554)$ & $(1.37)$ \\ \hline
Quadratic & $.487$ & $33.93$ \\ \cline{2-3}
& $(.640)$ & $(1.05)$ \\ \hline
Cubic & $-1.20$ & $32.27$ \\ \cline{2-3}
& $(.946)$ & $(.805)$ \\ \hline
\end{tabular}
\end{center}

We find statistically significant evidence of sensitivity to endogeneity for
the linear specification of demand but not for the quadratic or cubic. We
also find that the sensitivity estimates are quite small for all three
specifications. This absence of sensitivity of the AEV bound to endogeneity
suggests there is little need in this application to allow for price
endogeneity in the estimation of a lower bound on AEV.

\section{Endogenous Orthogonality Conditions}

There are many interesting economic and causal effects that depend on
functions satisfying endogenous orthogonality conditions where the function
of interest depends on variables that are not instruments. Such solutions to
orthogonality conditions come from first order conditions to economic choice
problems or define causal functions of interest. Objects of interest that
depend on such functions include policy and sensitivity effects like those
of Sections 2 and 3.

In this Section we derive the influence function for effects that depend on
the probability limit of a nonparametric instrumental variables (NPIV)
estimator like those in Newey and Powell (2003), Newey (1991), and Ai and
Chen (2003). We consider an estimator $\hat{\gamma}$ with a probability
limit $\gamma _{0}=\gamma (F_{0})$ that is the unique solution to
orthogonality conditions%
\begin{equation}
E[b(X)\rho (W,\gamma )]=0,\text{ }b\in \mathcal{B},\text{ }\gamma \in \Gamma
.  \label{orthog}
\end{equation}%
Here $\mathcal{B}$ is a linear set of possible instrumental variables $b(X)$
and $\gamma $ is restricted to a linear set $\Gamma $ similar to Section
3.1. We depart from Section 3.1 in allowing the unknown function $\gamma $
to depend on variables $Z$ that are different than the instruments $X.$ This
set up generalizes the conditional moment restrictions environment of Newey
and Powell (1989, 2003), Newey (1991), and Ai and Chen (2003) to
orthogonality conditions with linear restrictions on $\gamma .$

Restrictions on the structural functions and on the instrumental variables
are of interest to empirical researchers for at least two reasons. First
imposing correct restrictions on the structural function can improve
efficiency of the estimator and mitigate the well known ill-posed inverse
problem for NPIV that can lead to imprecise estimators. For example imposing
partially linear or additive structure on $\gamma $ can make estimators more
precise. Second imposing restrictions on the instrumental variables can help
reduce the well known Nagar (1959) instrumental variable bias. Such biases
are known to be important in empirical applications such as Angrist and
Kreuger (1991). By allowing such restrictions we provide the researcher with
more flexibility to choose a model that can lead to good inference
properties for policy or sensitivity analysis with endogeneity. We leave to
future work the application of the results of this Section to policy and
sensitivity analysis. We focus here on showing how Steps I and II can be
used to derive influence functions in complicated and important settings
which is a primary purpose of this paper.

\subsection{The Estimator}

We will derive influence functions for $\hat{\gamma}$ that is a first step
NPIV estimator based on the orthogonality conditions in equation (\ref%
{orthog}). Let $b^{K}(x)=(b_{1}(x),...,b_{K}(x))^{\prime }$ be the first $K$
elements of a sequence of instrumental variables. We assume that $b^{K}(X)$
spans $\mathcal{B}$ as $K$ grows meaning that any element of $\mathcal{B}$
can be approximated arbitrarily well by a linear combination of $b^{K}(X)$
for $K$ large enough. The NPIV estimator we consider is%
\begin{eqnarray}
\hat{\gamma} &=&\arg \min_{\gamma \in \Gamma _{n}}\hat{Q}(\gamma ),
\label{NPIV Est} \\
\hat{Q}(\gamma ) &=&\frac{1}{n}\sum_{i\text{.}=1}^{n}\rho (W_{i},\gamma
)b^{K}(X_{i})^{T}\left( \sum_{i=1}^{n}b^{K}(X_{i})b^{K}(X_{i})^{T}\right)
^{-}\sum_{i=1}^{n}b^{K}(X_{i})\rho (W_{i},\gamma ),  \notag
\end{eqnarray}%
where $\Gamma _{n}$ is a subset of $\Gamma $ and $A^{-}$ denotes a
generalized inverse of a matrix $A.$ For example, $\Gamma _{n}$ could be the
set of linear combinations of $L$ functions $p_{1}(z),...,p_{L}(z)$ where $%
p_{\ell }(\cdot )\in \Gamma $ for each $\ell $. We assume that a minimum
exists with probability approaching one, as could be guaranteed in some
settings using Chen and Pouzo (2015). This $\hat{\gamma}$ has the form of
NPIV given in Newey and Powell (1989, 2003), Newey (1991), Ai and Chen
(2003), and Darolles, Florens, and Renault (2011). We differ from this prior
work in allowing the instrumental variables to be restricted to the set $%
\mathcal{B}.$

The influence function for the object of interest will depend on the plim $%
\gamma _{\tau }$ of $\hat{\gamma}$ when the distribution of $W$ is $F_{\tau
}=(1-\tau )F_{0}+\tau H$. Since $\hat{\gamma}$ minimizes the sample
objective function $\hat{Q}(\gamma )$ the usual extremum estimator theory
(e.g. Amemiya, 1985), will imply that $\gamma _{\tau }$ is the minimum of
the plim $Q_{\tau }(\gamma )$ of $\hat{Q}(\gamma )$ when the distribution of 
$W$ is $F_{\tau }$. To describe $Q_{\tau }(\gamma )$ assume that $\mathcal{B}
$ does not depend on $\tau $, which can be shown to hold under regularity
conditions on $H$. Let $\pi _{\tau }(a(W)|X)$ denote the linear projection
of $a(W)$ on $\mathcal{B}$ when $W$ has CDF $F_{\tau },$ satisfying%
\begin{equation}
\pi _{\tau }(a(W)|X)\in \mathcal{B},\text{ }E_{\tau }[\{a(W)-\pi _{\tau
}(a(W)|X)\}b(X)]=0\text{ for all }b(X)\in \mathcal{B}  \label{proj}
\end{equation}%
Then it follows exactly as in Newey (1991) that for $K\longrightarrow \infty 
$ and $K/n\longrightarrow 0$, 
\begin{equation}
\text{plim}(\hat{Q}(\gamma ))=Q_{\tau }(\gamma ):=E_{\tau }[\left\{ \pi
_{\tau }(\rho (W,\gamma )|X)\right\} ^{2}].  \label{Qtau}
\end{equation}%
Intuitively, from standard regression results we see that $\hat{Q}(\gamma )$
is the sample average of squares of predicted values from the least squares
regression of $\rho (W_{i},\gamma )$ on $b^{K}(X_{i}),$ $(i=1,...,n).$ Then
by the law of large numbers, consistency of a sample regression for a
population regression, and the growth of $K$ it will follow that plim of $%
\hat{Q}(\gamma )$ will be the expected value of the square of the predicted
value from the population regression of $\rho (W,\gamma )$ on $\mathcal{B}$,
giving equation (\ref{Qtau}). It then follows by extremum estimator theory
and from $\Gamma _{n}$ assumed to approximate $\Gamma $ that%
\begin{equation*}
\text{plim}(\hat{\gamma})=\gamma _{\tau }:=\arg \min_{\gamma \in \Gamma
}Q_{\tau }(\gamma ).
\end{equation*}%
We will assume that $\gamma _{\tau }$ is unique, which could be shown to
hold under more primitive conditions in Chen and Pouzo (2015).

As in Section 3.2 the focus of this Section is deriving the FSIF $\phi
(w,\gamma ,\alpha )$ that satisfies $\partial E[m(W,\gamma _{\tau
})]/\partial \tau =\int \phi (w,\gamma _{0},\alpha _{0})H(dw).$ The first
order condition for $\gamma _{\tau }$ has a key role in deriving the FSIF.
To describe the first order condition let $\Delta \in \Gamma $ denote a
possible deviation of $\gamma $ away from $\gamma _{\tau }.$ Assume that
there is $v_{\rho \tau }(W)$ such that%
\begin{equation*}
\frac{\partial \pi _{\tau }(\rho (W,\gamma _{\tau }+\zeta \Delta )|X)}{%
\partial \zeta }=\pi _{\tau }(v_{\rho \tau }(W)\Delta (Z)|X).
\end{equation*}%
The calculus of variations, first order condition for the minimization of $%
Q(\gamma _{\tau }+\zeta \Delta )/2$ at $\zeta =0$ is 
\begin{eqnarray}
0 &=&\left. \frac{d}{d\zeta }E_{\tau }[\{\pi _{\tau }(\rho (W,\gamma _{\tau
}+\zeta \Delta )|X)\}^{2}]/2\right\vert _{\zeta =0}  \label{foc} \\
&=&E_{\tau }[\pi _{\tau }(\rho (W,\gamma _{\tau })|X)\frac{\partial \pi
_{\tau }(\rho (W,\gamma _{\tau }+\zeta \Delta )|X)}{\partial \zeta }]  \notag
\\
&=&E_{\tau }[\pi _{\tau }(\rho (W,\gamma _{\tau })|X)\pi _{\tau }(v_{\rho
\tau }(W)\Delta (Z)|X)]\text{ for all }\Delta \in \Gamma ,  \notag
\end{eqnarray}%
identically in $\tau $. This first order condition has a form analogous to
two-stage least squares, being orthogonality of the residual $\rho (W,\gamma
_{\tau })$ with instruments obtained by projecting the derivative of the
residual on the set of instrumental variables. We use this first order
condition and the orthogonality condition in equation (\ref{Qtau}) to
characterize the FSIF.

\subsection{The Adjustment Term}

Similarly to Section 3 the influence function of $\theta
(F)=E_{F}[m(W,\gamma (F))]$ will be the sum of $m(W,\gamma _{0})-\theta _{0}$
and the FSIF. We focus on derivation of the FSIF here. To characterize the
FSIF we proceed analogously to Section 3.2 by differentiating the first
order condition with respect to $\tau $ and applying the chain rule. For
notational simplicity let $\pi (A(W)|X)$ denote the projection of $A(W)$ on $%
\mathcal{B}$ for $\tau =0$. We carry out these calculations for the case
where $\pi (\rho (W,\gamma _{0})|X)=0$, where either the orthogonality
conditions are correctly specified or $\gamma _{0}$ is exactly identified so
that the plim of $\hat{\gamma}$ solves the orthogonality conditions (see
Chen and Santos, 2015, for exact identification). In Appendix B we derive
the FSIF under misspecification where $\pi (\rho (W,\gamma _{0})|X)\neq 0.$

Differentiating the identity of equation (\ref{foc}) with respect to $\tau $%
, using the third equality and $\pi (\rho (W,\gamma _{0})|X)=0,$ gives%
\begin{equation}
0=\frac{\partial }{\partial \tau }E[\pi _{\tau }(\rho (W,\gamma _{\tau
})|X)\pi (v_{\rho }(W)\Delta (Z)|X)]\text{ for all }\Delta \in \Gamma ,
\label{iv foc}
\end{equation}%
where $v_{\rho }(W)=v_{\rho 0}(W).$ Define the set $\mathcal{A}$ to be the
mean square closure of the set of $\pi (v_{\rho }(W)\Delta (Z)|X)$ for $%
\Delta \in \Gamma ,$ i.e. 
\begin{equation}
\mathcal{A=\{}\alpha (X):\text{for all }\varepsilon >0\text{ there is }%
\Delta (Z)\in \Gamma \text{ with }E[\{\alpha (X)-\pi (v_{\rho }(W)\Delta
(Z)|X)\}^{2}]<\varepsilon \},  \label{A def}
\end{equation}%
Then the first order condition in equation (\ref{foc}) becomes%
\begin{equation*}
0=\frac{\partial }{\partial \tau }E[\pi _{\tau }(\rho (W,\gamma _{\tau
})|X)\alpha (X)]\text{ for all }\alpha \in \mathcal{A}.
\end{equation*}

Next we use the orthogonality condition (\ref{proj}) for the projection.
Because $\mathcal{A}$ is a subset of $\mathcal{B}$ it follows that 
\begin{equation*}
E_{\tau }[\rho (W,\gamma _{\tau })\alpha (X)]=E_{\tau }[\pi _{\tau }(\rho
(W,\gamma _{\tau })|X)\alpha (X)]\text{ for all }\alpha \in \mathcal{A}
\end{equation*}%
identically in $\tau .$ Differentiating both sides with respect to $\tau $
and applying the chain rule gives%
\begin{equation*}
\frac{\partial }{\partial \tau }E_{\tau }[\rho (W,\gamma _{\tau })\alpha
(X)]=\frac{\partial }{\partial \tau }E_{\tau }[\pi (\rho (W,\gamma
_{0})|X)\alpha (X)]+\frac{\partial }{\partial \tau }E[\pi _{\tau }(\rho
(W,\gamma _{\tau })|X)\alpha (X)]=0,
\end{equation*}%
by $\pi (\rho (W,\gamma _{0})|X)=0$ and equation (\ref{iv foc}). Applying
the chain rule to the left-and side and solving then gives

\begin{equation}
-\frac{\partial }{\partial \tau }E[\rho (W,\gamma _{\tau })\alpha (X)]=\int
\alpha (x)\rho (w,\gamma _{0})H(dw)\text{ for all }\alpha \in \mathcal{A}.
\label{iv key}
\end{equation}%
Similarly to Section 3.1 the object being integrated on the right provides a
candidate for FSIF $\phi (w,\gamma ,\alpha )$. To find $\alpha _{0}(X)$ such
that equation (\ref{alpha ident}) is satisfied we impose the following
conditions.

\bigskip

\textsc{Assumption 3: }\textit{There exists }$v_{m}(Z)$\textit{\ such that }%
\begin{equation*}
\frac{\partial }{\partial \tau }E[m(W,\gamma _{\tau })]=\frac{\partial }{%
\partial \tau }E[v_{m}(Z)\gamma _{\tau }(Z)],\text{ }E[v_{m}(X)^{2}]<\infty .
\end{equation*}

\bigskip

This condition is analogous to Assumption 1 in specifying an expected
product form for $dE[m(W,\gamma _{\tau })]/d\tau ,$ and similarly will be
required for existence of the FSIF.

\bigskip

\textsc{Assumption 4: }\textit{\ There exists }$v_{\rho }(W)$\textit{\ such
that for all }$b\in B,$ 
\begin{equation*}
\frac{\partial }{\partial \tau }E[\rho (W,\gamma _{\tau })b(X)]=\frac{%
\partial }{\partial \tau }E[v_{\rho }(W)\gamma _{\tau }(Z)b(X)].
\end{equation*}

\bigskip

This condition is similar to Assumption 2 in specifying a derivative
condition involving the residual $\rho (W,\gamma )$ as a function of $\gamma 
$.

Unlike Section 3 the differentiability conditions in Assumptions 3 and 4 are
not sufficient to show that the FSIF has the form $\alpha (x)\rho (w,\gamma )
$ for some $\alpha _{0}(x).$ The presence of endogeneity, where $\gamma $
depends on variables different than the instrumental variables $X,$ creates
the need for a link between $v_{m}(Z),$ functions of $X,$ and $v_{\rho }(W)$%
. The following condition establishes the needed link. Let $\Pi
(d(W)|Z)=\arg \min_{\gamma \in \Gamma }E[\{d(W)-\gamma (Z)\}^{2}]$ denote
the least squares projection of a function $d(W)$ on $\Gamma .$

\bigskip

\textsc{Assumption 5: }\textit{There is }$b_{m}(X)\in \mathcal{B}$\textit{\
such that}%
\begin{equation*}
\Pi (v_{m}(Z)|Z)=-\Pi (v_{\rho }(W)b_{m}(X)|Z).
\end{equation*}

\bigskip

This condition requires that the projection of $v_{m}(Z)$ on $\Gamma $ must
be equal to the projection of $-v_{\rho }(W)b_{m}(X)$ on $\Gamma $ for some
instrumental variable $b_{m}(X).$ This condition is restrictive in a way
that is related to the Severini and Tripathi (2012) necessary conditions for
root-n consistent estimation as discussed in Example 6 to follow.

Assumptions 3-5 imply that the FSIF will have the form $\alpha (X)\rho
(W,\gamma )$ where $\alpha _{0}(X)$ is the least squares projection of $%
b_{m}(X)$ on $\mathcal{A}$. To see this note that by $\gamma _{\tau }\in
\Gamma $ and Assumption 5,%
\begin{eqnarray*}
E[v_{m}(Z)\gamma _{\tau }(Z)] &=&E[\Pi (v_{m}(Z)|Z)\gamma _{\tau
}(Z)]=-E[\Pi (v_{\rho }(W)b_{m}(X)|Z)\gamma _{\tau }(Z)] \\
&=&-E[v_{\rho }(W)b_{m}(X)\gamma _{\tau }(Z)]=-E[b_{m}(X)\pi (v_{\rho
}(W)\gamma _{\tau }(Z)|X)] \\
&=&-E[\alpha _{0}(X)\pi (v_{\rho }(W)\gamma _{\tau }(Z)|X)]=-E[\alpha
_{0}(X)v_{\rho }(W)\gamma _{\tau }(Z)],
\end{eqnarray*}%
for all $\tau $ where the fifth equality follows by $\pi (v_{\rho }(W)\gamma
_{\tau }(Z)|X)\in \mathcal{A}$. Then by Assumption 3 and 4 and
differentiating we have%
\begin{eqnarray*}
\frac{d}{d\tau }E[m(W,\gamma _{\tau })] &=&\frac{d}{d\tau }E[v_{m}(Z)\gamma
_{\tau }(Z)]=-\frac{d}{d\tau }E[\alpha _{0}(X)v_{\rho }(W)\gamma _{\tau }(Z)]
\\
&=&-\frac{d}{d\tau }E[\alpha _{0}(X)\rho (W,\gamma _{\tau })]=\int \alpha
_{0}(x)\rho (w,\gamma _{0})H(dw).
\end{eqnarray*}%
where the last equality follows from equation (\ref{iv key}). This equation
shows the following result:

\bigskip

\textsc{Proposition 3:} \textit{If Assumptions 3-5 are satisfied and }$\pi
(\rho _{0}(W,\gamma _{0})|X)=0$ \textit{then the FSIF is}%
\begin{equation*}
\phi (w,\gamma ,\alpha )=\alpha (x)\rho (w,\gamma ),
\end{equation*}%
\textit{where }$\alpha _{0}(X)$\textit{\ is the least squares projection of }%
$b_{m}(X)$\textit{\ on }$A$\textit{\ satisfying}%
\begin{equation*}
\alpha _{0}(X)=\arg \min_{\alpha \in \mathcal{A}}E[\{b_{m}(X)-\alpha
(X)\}^{2}].
\end{equation*}

\bigskip

The derivation of Proposition 3 is more complicated than Proposition 1
because of endogeneity and the link condition in Assumption 5. The function $%
\alpha _{0}(X)$ quantifies how the instrumental variables affect the FSIF.
It is constrained to be an element of $\mathcal{A}$ because NPIV projects
functions of $Z$ on the set of instrumental variables $\mathcal{B}$, just as
parametric two-stage least square does. When multiple sets of orthogonality
conditions are available, e.g. as could be the case if $E[\rho (W,\gamma
_{0})|X]=0$, $\alpha _{0}(X)$ can vary with $\mathcal{B}$. This effect of
the choice of $\mathcal{B}$ on the influence function is analogous to
parametric instrumental variables estimation, where the influence function
can vary with the choice of linear combination of instrumental variables.

\bigskip

\textsc{Example 5:} \textit{Additive Structural Functions and Instruments; }%
We consider NPIV where $\gamma (Z)=\gamma _{1}(Z_{1})+\gamma (Z_{2})$ is
restricted to be additive in distinct components $Z_{1}$ and $Z_{2}$ of $%
Z=(Z_{1},Z_{2}).$ Such a restriction can reduce the severity of the
ill-posed inverse problem. The instrumental variables $%
b(X)=b_{1}(X_{1})+b_{2}(X_{2})$ are also restricted to be additive in
distinct components of $X_{1}$ and $X_{2}$ of $X.$ Such a restriction can
identify the additive components $\gamma _{1}(Z_{1})$ and $\gamma
_{2}(Z_{2}) $ while limiting the number of instrumental variables to reduce
the Nagar (1959) bias of instrumental variables estimators. Here $\Gamma $
and $\mathcal{B}$ are mean square closures of sets of additive functions. It
will be convenient here to just refer to additive functions rather the mean
square closures of sets of functions, though not every function in the
closure need be additive.

One thing of note about the FSIF here is that $\alpha _{0}(X)$ is in $%
\mathcal{B}$ and so it is an additive function of $X_{1}$ and $X_{2}$. The
form of $\alpha _{0}(X)$ will be determined by the form of $v_{m}(Z)$ and $%
v_{\rho }(W)$ and the link condition of Assumption 5. Here $\Pi (A(W)|Z)$ is
the projection on (the mean square closure of) additive functions. Also the
elements of $\mathcal{B}$ are (in the closure of) additive functions.
Suppose that the residual is linear with%
\begin{equation*}
\rho (W,\gamma )=Y-\gamma _{1}(Z_{1})-\gamma _{2}(Z_{2}).
\end{equation*}%
Then $v_{\rho }(W)=-1$ so that Assumption 5 is existence of $b_{m}\in 
\mathcal{B}$ with%
\begin{equation*}
\Pi (v_{m}(Z)|Z)=\Pi (b_{m}(X)|Z).
\end{equation*}%
This requires that the projection of $v_{m}(Z)$ on additive functions of $%
Z_{1}$ and $Z_{2}$ must be equal to the projection of an additive function
of $X_{1}$ and $X_{2}$ on additive functions of $Z_{1}$ and $Z_{2}.$ For
example if $Z_{1}$ is a scalar and $m(w,\gamma )=\omega (z_{1})\partial
\gamma _{1}(z_{1})/\partial z_{1}$ then as in Example 1,%
\begin{equation*}
v_{m}(Z)=-\frac{1}{f_{0}(Z)}\frac{\partial \{w(Z_{1})f_{0}(Z)\}}{\partial
z_{1}}=-\frac{\partial w(Z_{1})}{\partial z_{1}}-w(Z_{1})\frac{\partial
f_{0}(Z)/\partial z_{1}}{f_{0}(Z)}.
\end{equation*}%
Here it would suffice for Assumption 5 that there $b^{I}(X_{1})$ and $%
b^{II}(X)=b_{1}^{II}(X_{1})+b_{2}^{II}(X_{2})$ such that%
\begin{equation}
-\frac{\partial w(Z_{1})}{\partial z_{1}}=E[b^{II}(X_{1})|Z_{1}]\text{, }\Pi
(w(Z_{1})\frac{\partial f_{0}(Z)/\partial z_{1}}{f_{0}(Z)}|Z)=\Pi
(b^{II}(X)|Z)  \label{endog der}
\end{equation}

For quantile orthogonality conditions where $\rho (W,\gamma )=p-1(Y<\gamma
(Z))$, it follows similarly to Section 3 that%
\begin{equation*}
v_{\rho }(W)=f(\gamma _{0}(Z)|Z,X),
\end{equation*}%
where $f(Y|Z,X)$ is the pdf of $Y$ conditional on $Z$ and $X.$ Assumption 5
is then existence of $b_{m}\in \mathcal{B}$ with%
\begin{equation*}
\Pi (v_{m}(Z)|Z)=\Pi (f(\gamma _{0}(Z)|Z,X)b_{m}(X)|Z).
\end{equation*}%
This requires that the projection of $v_{m}(Z)$ on additive functions of $%
Z_{1}$ and $Z_{2}$ must be equal to the projection of a weighted additive
function of $X_{1}$ and $X_{2}$ on additive functions of $Z_{1}$ and $Z_{2}.$
This condition also restricts $v_{m}(Z)$ to be such that its projection on $%
\Gamma $ is equal to projection of a function of $Z$ and $X$ on $\Gamma $ as
further discussed in Example 7 to follow.

To help relate Proposition 3 to prior work we consider a simple example of
an object of interest for conditional moment restrictions.

\bigskip

\textsc{Example 6:} \textit{Linear Function of a Linear Structural Equation; 
}A relatively simple example has $m(W,\gamma )=v_{m}(Z)\gamma (Z)$ for a $%
v_{m}(Z)$ with $E[v_{m}(Z)^{2}]<\infty ,$ $\rho (w,\gamma )=y-\gamma (z),$
and $\Gamma $ and $\mathcal{B}$ are unrestricted, so that the orthogonality
condition of equation (\ref{orthog}) is 
\begin{equation*}
Y=\gamma _{0}(Z)+\varepsilon ,\text{ }E[\varepsilon |X]=0.
\end{equation*}%
This is a linear NPIV equation. Assumptions 3 and 4 are satisfied with $%
v_{m}(Z)$ as given in this example and $v_{\rho }(W)=-1.$ Then Assumption 5
is existence of $b_{m}(X)$ such that%
\begin{equation}
v_{m}(Z)=E[b_{m}(X)|Z].  \label{sevtrip}
\end{equation}%
Also $\mathcal{A}$ is the mean square closure of $E[\Delta (Z)|X]$ over all $%
\Delta (Z)$ with finite second moment and $\alpha _{0}(X)$ is the projection
of $b_{m}(X)$ on $\mathcal{A}$. The FSIF is then%
\begin{equation}
\phi (W,\gamma _{0},\alpha _{0})=\alpha _{0}(X)\{Y-\gamma _{0}(Z)\}.
\label{npiv adj}
\end{equation}

It is interesting to note that existence of a solution $b_{m}(X)$ to
equation (\ref{sevtrip}) is the necessary condition of Severini and Tripathi
(2012) for existence of a root-n consistent estimator of $\theta
_{0}=E[v_{m}(Z)\gamma _{0}(Z)].$ This condition is restrictive in imposing
that coefficients in a singular value expansion of $b_{m}(X)$ must decline
at certain rates. This example shows the precise relationship of that
necessary condition to the $\alpha _{0}(X)$ in the FSIF. The $\alpha _{0}(X)$
is the projection of $b_{m}(X)$ on $\mathcal{A}$.

The formula for the FSIF given here is related to a prior influence function
formula given in Ai and Chen (2007, p. 40) for conditional moment
restrictions. In the notation here the Ai and Chen (2007) formula is 
\begin{equation}
\phi (W,\gamma _{0},\alpha _{0})=E[v^{\ast }(Z)|X]\{Y-\gamma _{0}(Z)\},
\label{AiChen adj}
\end{equation}%
where $v^{\ast }(Z)$ is a Riesz representer in an extended Hilbert space
described in Ai and Chen (2003, 2007). Equations (\ref{npiv adj}) and (\ref%
{AiChen adj}) coincide for $\alpha _{0}(X)=E[v^{\ast }(Z)|X]$. Equation (\ref%
{npiv adj}) is more explicit in giving the precise relationship between $%
\alpha _{0}(X)$ and the $b_{m}(X)$ of the Severini and Tripathi (2012)
necessary condition. Also Proposition 3 allows orthogonality conditions that
are more general than conditional moment restrictions. Interesting and
useful Hilbert space characterizations of the FSIF in Proposition 3 could be
obtained as in Chen and Liao (2015) and/or Chen and Pouzo (2015) by
extending their results for conditional moment restrictions to orthogonality
conditions. The more explicit formula in Proposition 3 may prove useful for
policy and sensitivity analysis and the construction of orthogonal moment
functions.

The NPIV objective function in equation (\ref{NPIV Est}) can be modified to
allow a weighted second moment matrix in the middle as in Ai and Chen (2003)
where $\sum_{i=1}^{n}b^{K}(X_{i})b^{K}(X_{i})^{T}$ is replaced by $%
\sum_{i=1}^{n}\omega (X_{i})b^{K}(X_{i})b^{K}(X_{i})^{T}$ for $\omega
(X_{i})>0.$ Such a modification with $\omega (X_{i})=Var(\rho (W_{i},\gamma
)|X_{i})$ would lead to improved asymptotic efficiency of $\hat{\theta}$ if $%
\gamma $ were a finite dimensional parameter vector and $m(W,\gamma )$ did
not depend on $W$. Proposition 3 can be modified in a straightforward way to
allow for the presence of such a $\omega (X_{i})$ by replacing $\rho
(W,\gamma )$ with $\omega (X)^{-1}\rho (W,\gamma )$ and $E_{\tau }[\cdot ]$
with the weighted expectation $E_{\tau }[\omega (X)(\cdot )],$ including in
the projection $\pi .$ Further details are beyond the scope of this paper.

\section{Extensions and Conclusions}

It is straightforward to extend the results we have given to objects that
depend on multiple nonparametric estimators. As discussed in Newey (1994)
such objects will have a separate FSIF for each nonparametric estimator and
the overall FSIF will be the sum of the separate adjustment terms. Also,
each separate FSIF can be computed from varying one nonparametric estimator
while holding the others fixed at their limit. It is also straightforward to
extend the results to objects of interest that maximize objective functions
other than that for GMM. This extension is described in Appendix B.

This paper gives explicit influence function formulae for first steps that
satisfy exogenous or endogenous orthogonality conditions. It is shown how
such formulae are useful for characterizing local policy effects of
structural changes, quantifying sensitivity of semiparametric estimators,
and constructing orthogonal moment functions. Those results are used to
generalize the omitted variable bias formula for regression to obtain the
local effect of misspecification on policies and estimators that depend on
solutions to exogenous orthogonality conditions. This analysis is applied to
a gasoline demand data set where we find no evidence that average equivalent
variation bounds are sensitive to endogeneity.\pagebreak 

\begin{center}
\textsc{Appendix A: Validity of the influence function calculation}
\end{center}

In this Appendix we show validity of Steps I and II of the influence
function calculation. Step I requires differentiability of $\theta (F_{\tau
})$ and the formula%
\begin{equation}
\frac{d\theta (F_{\tau })}{d\tau }=\int \psi (w)H(dw),\text{ }E[\psi (W)]=0,%
\text{ }E[\psi (W)^{2}]<\infty .  \label{Inf deriv}
\end{equation}%
Step II) requires that evaluating the derivative at a point mass gives the
influence function. We justify Step II) as a limit as $H$ approaches a point
mass similarly to Lebesgue differentiation from analysis. Lebesgue
differentiation shows that the limit of an integral of a function over an
interval divided by the length of the interval converges almost surely to
the value of the function at a point as the interval collapses on that
point. We give regularity conditions and classes of continuous, smooth
probability distributions where expectation of the influence function
converges to its value at a point as the probability distribution collapses
on the point.

The fundamental starting point for the influence function calculation is
that the estimator is asymptotically linear with an influence function, i.e.
that it satisfies%
\begin{equation*}
\sqrt{n}(\hat{\theta}-\theta _{0})=\frac{1}{\sqrt{n}}\sum_{i=1}^{n}\psi
(W_{i})+o_{p}(1),\text{ }E[\psi (W)]=0,\text{ }E[\psi (W)^{T}\psi
(W)]<\infty .
\end{equation*}%
We take a modern, high level approach to regularity conditions in assuming
that the estimator is locally regular for a set of alternative distributions 
$H$ that can approximate a point mass.

\bigskip

\textsc{Definition A1:} $\hat{\theta}$ \textit{is locally regular for }$%
F_{\tau }$\textit{\ if there is a fixed random variable }$Y$\textit{\ such
that for any }$\tau _{n}=O(1/\sqrt{n})$\textit{\ and }$W_{1},...,W_{n}$%
\textit{\ i.i.d. with distribution }$F_{\tau _{n}},$%
\begin{equation*}
\sqrt{n}[\hat{\theta}-\theta (F_{\tau _{n}})]\overset{d}{\longrightarrow }Y.
\end{equation*}

\bigskip

This local regularity condition is familiar from the efficient estimation
literature. Local regularity of $\hat{\theta}$ is not a primitive condition
but it is plausible when $F_{0}$ satisfies conditions for existence of $%
\theta (F)$ and $H$ is well behaved relative to $F_{0}.$ For example $F_{0}$
could satisfy regularity conditions like some random variables being
continuously distributed and expectations of certain functions existing and $%
H$ could be a uniformly bounded, very smooth deviation from $F_{0}.$ In such
settings it is plausible many estimators $\hat{\theta}$ would be locally
regular. We construct such $H$ in this Appendix so that local regularity is
plausibly satisfied for many semiparametric estimators $\hat{\theta}$.

We consider a sequence $(H_{w}^{j})_{j=1}^{\infty }$ taking the form%
\begin{equation}
H_{w}^{j}(\tilde{w})=E[1(W\leq \tilde{w})\delta _{w}^{j}(W)],  \label{G def}
\end{equation}%
where for each $j$ the random variable $\delta _{w}^{j}(W)$ is bounded with $%
E[\delta _{w}^{j}(W)]=1$. In $H_{w}^{j}(\tilde{w})$ the variable $\tilde{w}$
represents a possible value of the random variable $W.$ As we will discuss
this $H_{w}^{j}(\tilde{w})$ will have the needed properties when $\delta
_{w}^{j}(W)$ is chosen appropriately. In particular the support of $%
H_{w}^{j}(\tilde{w})$ will approach $\{w\}$ as the support of $\delta
_{w}^{j}(w)$ does. Throughout we will assume that $w$ is a vector of real
numbers of fixed dimension $r.$ We impose the following properties:

\bigskip

\textsc{Assumption A1:} $F_{0}$\textit{\ is absolutely continuous with
respect to a measure }$\mu $\textit{\ on }$%
\mathbb{R}
^{r}$\textit{\ with pdf }$f_{0}(w),$ $\delta _{w}^{j}(W)$\textit{\ is not
constant, bounded, and }$E[\delta _{w}^{j}(W)]=1$\textit{.}

\bigskip

By $\delta _{w}^{j}(W)$ bounded $F_{\tau }^{j}=(1-\tau )F_{0}+\tau H_{w}^{j}$
will be a CDF for small enough $\tau $ with pdf with respect to $\mu $ given
by 
\begin{equation}
f_{\tau }(\tilde{w})=f_{0}(\tilde{w})[1-\tau +\tau \delta _{w}^{j}(\tilde{w}%
)]=f_{0}(\tilde{w})[1+\tau S(\tilde{w})],S(\tilde{w})=\delta _{w}^{j}(\tilde{%
w})-1,  \label{submodel}
\end{equation}%
where we suppress the $j$ superscript and $w$ subscript on $f_{\tau }(\tilde{%
w})$ and $S(\tilde{w})$ for notational convenience. Note that by $S(\tilde{w}%
)$ bounded there is $C$ such that for small enough $\tau ,$%
\begin{equation}
(1-\tau )f_{0}/C\leq f_{\tau }\leq Cf_{0},  \label{abscont}
\end{equation}%
so that $f_{\tau }$ and $f_{0}$ will be absolutely continuous with respect
to each other. Thus, variables that are continuously distributed under $%
F_{0} $ will also be continuously distributed under $F_{\tau }^{j}$. Also
objects that have expectation close to zero for $F_{0}$ will also have
expectation close to zero under $F_{\tau }^{j}$ and vice versa. If $\theta
(F)$ being well defined depends on existence of derivatives of the pdf for $%
F $ then that restriction can be imposed by choosing $\delta _{w}^{j}(\tilde{%
w})$ so its derivatives exist. In these ways we can choose $\delta
_{w}^{j}(w)$ so that $f_{\tau }(\tilde{w})$ satisfies the restrictions
needed for $\theta (F_{\tau }^{j})$ to be well defined$.$

We assume that the sequence $(\delta _{w}^{j})_{j=1}^{\infty }$ satisfies a
condition leading to%
\begin{equation}
\lim_{j\longrightarrow \infty }\int \psi (\tilde{w})H_{w}^{j}(d\tilde{w}%
)\longrightarrow \psi (w),  \label{inf lim}
\end{equation}%
thus justifying Step II of the influence function calculation. Define a
function $a(\tilde{w})$ to be almost surely continuous at $w$ in $\mu $ if
for any $\varepsilon >0$ there is a neighborhood $N$ of $w$ and a subset $%
N_{\mu }$ of $N$ such that $\mu (N_{\mu })=\mu (N)$ and $|a(\tilde{w}%
)-a(w)|<\varepsilon $ for all $\tilde{w}\in N_{\mu }.$

\bigskip

\textsc{Assumption A2:} \textit{If }$a(\tilde{w})$\textit{\ is }$\mu $ 
\textit{almost surely} \textit{continuous at }$w$\textit{\ and }$%
E[a(W)^{2}]<\infty $\textit{\ then }$\delta _{w}^{j}(W)$ \textit{satisfies }$%
\lim_{j\longrightarrow \infty }E[a(W)\delta _{w}^{j}(W)]=a(w).$

\bigskip

This Assumption will be sufficient for equation (\ref{inf lim}). There are a
variety of ways that $\delta _{w}^{j}(W)$ can be chosen so that Assumption 2
will be satisfied. The basic idea is to consider $w$ where $f_{0}(\tilde{w})$
is bounded away from zero on a neighborhood of $w$ in the support of $W$ and
choose $\delta _{w}^{j}(\tilde{w})=g_{w}^{j}(\tilde{w})/f_{0}(\tilde{w})$
where $g_{w}^{j}(\tilde{w})$ is a bounded pdf and the support of $g_{w}^{j}(%
\tilde{w})$ to converge to $\,\{w\}.$ In the Appendix D we will choose $%
g_{w}^{j}(\tilde{w})$ in a way that is helpful for endogenous orthogonality
conditions. Another choice of $g_{w}^{j}(\tilde{w})$ that will lead to
equation (\ref{inf lim}) in many cases can be based on a nonnegative kernel $%
K(u)$ with bounded support $S$, as in the following result.

\bigskip

\textsc{Lemma A1: }\textit{If i) }$K(u)\geq 0$\textit{, }$\int K(u)du=1,$%
\textit{\ and }$K(u)$\textit{\ has bounded support \ }$S$\textit{; ii) there
is a neighborhood }$N$\textit{\ of }$w$\textit{\ and }$C>0$\textit{\ such
that }$f_{0}(\tilde{w})\geq C$\textit{\ almost surely }$\mu $\textit{\ for }$%
\tilde{w}\in N$\textit{; iii) }$\mu (w+\sigma S)>0$\textit{\ for all }$%
\sigma >0;$\textit{\ then for any }$(\sigma (j))_{j=1}^{\infty }$\textit{\
with }$\sigma (j)>0,$\textit{\ }$\sigma (j)\longrightarrow 0,$\textit{\ and }%
$w+\sigma (j)S\subseteq N$\textit{\ for all }$\sigma (j)$\textit{,
Assumptions 1 and 2 are satisfied for}%
\begin{equation*}
\delta _{w}^{j}(W)=f_{0}(W)^{-1}[\int 1(\tilde{w}\in w+\sigma (j)S)\sigma
(j)^{-r}K\left( \frac{\tilde{w}-w}{\sigma (j)}\right) \mu (d\tilde{w}%
)]^{-1}\sigma (j)^{-r}K\left( \frac{W-w}{\sigma (j)}\right) .
\end{equation*}

Note that if W has the Lebesgue density $f_{0}$, then the expression for $%
\delta _{w}^{j}$ simplifies to%
\begin{equation*}
\delta _{w}^{j}(\tilde{w})=f_{0}(\tilde{w})^{-1}\sigma (j)^{-r}K\left( \frac{%
\tilde{w}-w}{\sigma (j)}\right) .
\end{equation*}

\textsc{Proof:} Note that $\ $%
\begin{equation*}
\int 1(\tilde{w}\in W+\sigma (j)S)\sigma (j)^{-r}K\left( \frac{\tilde{w}-W}{%
\sigma (j)}\right) \mu (d\tilde{w})>0
\end{equation*}%
by i) and iii). Also, $K\left( (W-w)/\sigma (j)\right) $ is nonzero only on
a subset of $N$ so that $\delta _{w}^{j}(W)$ is bounded by i and ii). In
addition $E[\delta _{w}^{j}(W)]=1$ by construction.

Suppose $a(W)$ has finite second moment and is continuous at $w$ a.s. $\mu .$
Then for any $\varepsilon >0$ there is $j_{\varepsilon }$ large enough such
that for $j\geq j_{\varepsilon },$%
\begin{equation*}
a(w)-\varepsilon \leq a(W)\leq a(w)+\varepsilon
\end{equation*}%
a.s. $\mu $ for $W\in w+\sigma (j)S.$ Since $\delta _{w}^{j}(W)$ is
nonnegative and nonzero only on $W\in w+\sigma (j)S$ we have%
\begin{equation*}
a(w)-\varepsilon =E[\{a(w)-\varepsilon \}\delta _{w}^{j}(W)]\leq
E[a(W)\delta _{w}^{j}(W)]\leq E[\{a(w)+\varepsilon \}\delta
_{w}^{j}(W)]=a(w)+\varepsilon ,
\end{equation*}%
for all $j\geq j_{\varepsilon }.$ The conclusion follows by $\varepsilon $
being any positive number. $Q.E.D.$

\bigskip

The choice of $\delta _{w}^{j}(W)$ in Lemma A1 is simply a device to help
the limit of the Gateaux derivative exist under as general conditions as
possible. The limit, and hence the influence function, does not depend on
the kernel. Also, we could replace the continuity of $a(\tilde{w})$ at $w$
in Assumption 1 with other conditions that are sufficient for equation (\ref%
{inf lim}) on a set of $w$ with probability one under $F_{0}$. Equation (\ref%
{inf lim}) is analogous to the Lebesgue differentiation theorem that is
known to hold under quite general conditions on $a(\tilde{w})$. For example,
for the $\delta _{w}^{j}(w)$ of Lemma A1 equation (\ref{inf lim}) can be
shown to hold for any measurable $a(\tilde{w})$ if $\mu $ is the sum of
Lebesgue measure and a measure with a finite number of atoms. We use the
continuity condition of Assumption A1 because it is relatively simple to
state and because many influence functions will be $\mu $ almost sure
continuous on a set of $w$ that has probability one.

The next result shows that the influence function formula (\ref{inf lim}) is
valid for $H_{w}^{j}$ as specified in equation (\ref{G def}).

\bigskip

\textsc{Theorem A2:} \textit{If Assumptions A1 and A2 are satisfied, }$\hat{%
\theta}$\textit{\ is asymptotically linear with influence function }$\psi (%
\tilde{w}),$ $\hat{\theta}$ \textit{is locally regular for }$F_{\tau }^{j}(%
\tilde{w})=(1-\tau )F_{0}(\tilde{w})+\tau H_{w}^{j}(\tilde{w})$ \textit{for
each integer }$j$ \textit{and} $H_{w}^{j}(\tilde{w})=E[1(W\leq \tilde{w}%
)\delta ^{j}(W)],$\textit{\ and }$\psi (\tilde{w})$\textit{\ is }$\mu $ 
\textit{almost surely continuous at }$w,$ \textit{then }$d\theta (F_{\tau
}^{j})/d\tau $ \textit{exists, }$d\theta (F_{\tau }^{j})/d\tau =\int \psi (%
\tilde{w})H_{w}^{j}(d\tilde{w}),$\textit{\ and equation (\ref{inf lim}) is
satisfied.}

\bigskip

\textsc{Proof:} \ By $S(\tilde{w})=\delta _{w}^{j}(\tilde{w})-1$ bounded
there is an open set $T$ containing zero such that for all $\tau \in T,$ $%
1+\tau S(\tilde{w})$ is positive, bounded away from zero, and $f_{\tau }(%
\tilde{w})^{1/2}=f_{0}(\tilde{w})^{1/2}[1+\tau S(\tilde{w})]^{1/2}$ is
continuously differentiable in $\tau $ with 
\begin{equation*}
s_{\tau }(\tilde{w})=\frac{d}{d\tau }f_{0}(\tilde{w})^{1/2}[1+\tau S(\tilde{w%
})]^{1/2}=\frac{1}{2}\frac{f_{0}(\tilde{w})^{1/2}S(\tilde{w})}{[1+\tau S(%
\tilde{w})]^{1/2}}\leq Cf_{0}(\tilde{w})^{1/2}S(\tilde{w}).
\end{equation*}%
By $S(\tilde{w})$ bounded, $\int \left[ Cf_{0}(\tilde{w})^{1/2}S(\tilde{w})%
\right] ^{2}d\mu <\infty .$ Then by the dominated convergence theorem $f_{0}(%
\tilde{w})^{1/2}[1+\tau S(\tilde{w})]^{1/2}$ is mean-square differentiable
and $I(\tau )=\int s_{\tau }(\tilde{w})^{2}d\mu $ is continuous in $\tau $
on a neighborhood of zero. By Assumption 1 $S(W)$ is not zero so that $%
I(\tau )>0$. Then by Theorem 7.2 and Example 6.5 of Van der Vaart (1998) it
follows that for any $\tau _{n}=O(1/\sqrt{n})$ a vector of $n$ observations $%
(W_{1},...,W_{n})$ that is i.i.d. with pdf $f_{\tau _{n}}(\tilde{w})$ is
contiguous to $(W_{1},...,W_{n})$ that is i.i.d. with pdf $f_{0}(\tilde{w})$%
. Therefore, 
\begin{equation*}
\sqrt{n}(\hat{\theta}-\theta _{0})=\frac{1}{\sqrt{n}}\sum_{i=1}^{n}\psi
(W_{i})+o_{p}(1)
\end{equation*}%
holds when $(W_{1},...,W_{n})$ are i.i.d. with pdf $f_{\tau _{n}}(\tilde{w})$%
.

Next define $\mu _{w}^{j}=E[\psi (W)S(W)]=E[\psi (W)\delta _{w}^{j}(W)].$
Then by $E[\psi (W)]=0$, 
\begin{equation*}
E_{\tau }[\psi (W)]=\tau \mu _{w}^{j}.
\end{equation*}%
Suppose $(W_{1},...,W_{n})$ are i.i.d. with pdf $f_{\tau _{n}}(\tilde{w}).$
Let $\theta (\tau )=\theta ((1-\tau )F_{0}+\tau G_{w}^{j}),$ $\theta
_{n}=\theta (\tau _{n})$, and $\breve{\psi}_{n}(W)=\psi (W)-\tau _{n}\mu
_{w}^{j}.$ Adding and subtracting terms, 
\begin{eqnarray*}
\sqrt{n}\left( \hat{\theta}-\theta _{n}\right) &=&\sqrt{n}(\hat{\theta}%
-\theta _{0})-\sqrt{n}(\theta _{n}-\theta _{0})=\frac{1}{\sqrt{n}}%
\sum_{i=1}^{n}\psi (W_{i})+o_{p}(1)-\sqrt{n}(\theta _{n}-\theta _{0}) \\
&=&\frac{1}{\sqrt{n}}\sum_{i=1}^{n}\breve{\psi}_{n}(W_{i})+o_{p}(1)+\sqrt{n}%
\tau _{n}\mu _{w}^{j}-\sqrt{n}(\theta _{n}-\theta _{0}).
\end{eqnarray*}%
Note that $E_{\tau _{n}}[\breve{\psi}_{n}(W)]=0$. Also, by $\tau _{n}$
bounded, 
\begin{eqnarray*}
E_{\tau }[1(\left\Vert \breve{\psi}_{n}(W)\right\Vert &\geq &M)\left\Vert 
\breve{\psi}_{n}(W)\right\Vert ^{2}]\leq CE[1(\left\Vert \breve{\psi}%
_{n}(W)\right\Vert \geq M)\left\Vert \breve{\psi}_{n}(W)\right\Vert ^{2}] \\
&\leq &CE[1(\left\Vert \breve{\psi}_{n}(W)\right\Vert \geq M)(\left\Vert
\psi (W)\right\Vert ^{2}+C)] \\
&\leq &CE[1(\left\Vert \psi (W)\right\Vert \geq M-C)(\left\Vert \psi
(W)\right\Vert ^{2}+C)]\longrightarrow 0,
\end{eqnarray*}%
as $M\longrightarrow \infty $, so the Lindbergh-Feller condition for a
central limit theorem is satisfied. Furthermore, it follows by similar
calculations that $E_{\tau _{n}}[\breve{\psi}_{n}(W)\breve{\psi}%
_{n}(W)^{T}]\longrightarrow V.$ Therefore, by the Lindbergh-Feller central
limit theorem, $\sum_{i=1}^{n}\breve{\psi}_{n}(W_{i})/\sqrt{n}\overset{d}{%
\longrightarrow }N(0,V)$. By local regularity$\sqrt{n}(\hat{\theta}-\theta
_{n})\overset{d}{\longrightarrow }N(0,V)$ implying that%
\begin{equation}
\sqrt{n}\tau _{n}\mu _{w}^{j}-\sqrt{n}(\theta _{n}-\theta
_{0})\longrightarrow 0.  \label{mean converge}
\end{equation}

Next, we follow the proof of Theorem 2.1 of Van der Vaart (1991). The above
argument shows that local regularity implies that eq. (\ref{mean converge})
holds for all $\tau _{n}=O(1/\sqrt{n}).$ Consider any sequence $%
r_{m}\longrightarrow 0$. Let $n_{m}$ be the subsequence such that 
\begin{equation*}
(1+n_{m})^{-1/2}<r_{m}\leq n_{m}^{-1/2}.
\end{equation*}%
Let $\tau _{n}=r_{m}$ for $n=n_{m}$ and $\tau _{n}=n^{-1/2}$ for $n\notin
\{n_{1},n_{2},...\}.$ By construction, $\tau _{n}=O(1/\sqrt{n}),$ so that eq
(\ref{mean converge}) holds. Therefore it also holds along the subsequence $%
n_{m}$, so that 
\begin{equation*}
\sqrt{n_{m}}r_{m}\left\{ \mu _{z}^{j}-\frac{\theta (r_{m})-\theta _{0}}{r_{m}%
}\right\} =\sqrt{n_{m}}r_{m}\mu _{z}^{j}-\sqrt{n_{m}}[\theta (r_{m})-\theta
_{0}]\longrightarrow 0.
\end{equation*}%
By construction $\sqrt{n_{m}}r_{m}$ is bounded away from zero, so that $\mu
_{z}^{h}-\left[ \theta (r_{m})-\theta _{0}\right] /r_{m}\longrightarrow 0$.
Since $r_{m}$ is any sequence converging to zero it follows that $\theta
(\tau )$ is differentiable at $\tau =0$ with derivative $\mu _{z}^{j}$. The
conclusion then follows by Assumption 2. $Q.E.D.$

\bigskip

Let $H_{w}^{\infty }$ be the CDF with $\Pr (W=w)=1.$ Theorem A2 gives
sufficient conditions for equation (\ref{inf lim}) which is 
\begin{equation*}
\psi (w)=\int \psi (\tilde{w})H_{w}^{\infty }(d\tilde{w})=\lim_{H_{w}^{j}%
\longrightarrow H_{w}^{\infty }}\int \psi (\tilde{w})H_{w}^{j}(d\tilde{w}),
\end{equation*}%
where the first equality holds by definition of $H_{w}^{\infty }$. The
second equality states that $\psi (w)$ is the Lebesgue derivative of $\int
\psi (\tilde{w})H(d\tilde{w})$ based on the regularity conditions of
Assumptions A1 and A2 and the sequences of functions detailed there. This
Lebesgue differentiation conclusion justifies Step II of the Gateaux
derivative calculation as simply evaluating the Lebesgue derivative at a
point. This evaluation will be valid with probability one under Assumptions
A1 and A2.

We emphasize that the purpose of Theorem 2 is quite different than the
results of Bickel, Klaasen, Ritov, and Wellner (1993), Van der Vaart (1991)
and other important contributions to the semiparametric efficiency
literature. Here $\theta \left( F\right) $ is not a parameter of some
semiparametric model. Instead $\theta (F)$ is associated with an \textit{%
estimator }$\hat{\theta}$, being the probability limit of that estimator
when $F$ is a distribution that is unrestricted except for regularity
conditions, as in Newey (1994). Our goal is to use $\theta (F)$ to calculate
the influence function of $\hat{\theta}$ under the assumption that $\hat{%
\theta}$ is asymptotically linear. The purpose of Theorem A2 is to justify
Steps I and II as a way to do that calculation. In contrast, the goal of the
semiparametric efficiency literature is to find the efficient influence
function for a parameter of interest when $F$ belongs to a family of
distributions.

To highlight this contrast, note that the Gateaux derivative limit
calculation can be applied to obtain the influence function under
misspecification while efficient influence function calculations generally
impose correct specification. Indeed, the definition of $\theta (F)$ \textit{%
requires }that misspecification be allowed for, because $\theta (F)$ is
limit of the estimator $\theta $ under all distributions $F$ that are
unrestricted except for regularity condition. Of course correct
specification may lead to simplifications in the form of the influence
function. Such simplifications will be incorporated automatically when the
Gateaux derivative limit is taken at an $F_{0}$ that satisfies model
restrictions.

Theorem 2 is like Van der Vaart (1991, Theorem 2.1) in having
differentiability of $\theta (F_{\tau })$ as a conclusion. It differs in
restricting the paths to have the form $(1-\tau )F_{0}+\tau H_{w}^{j}$. Such
a restriction on the paths actually weakens the local regularity hypothesis
because $\theta $ only has to be locally regular for a particular kind of
path rather than the general class of paths in Van der Vaart (1991). We note
that this result allows for the distribution of $W$ to have discrete
components because the dominating measure $\mu $ may have atoms.

The weak nature of the local regularity condition highlights the strength of
the asymptotic linearity hypothesis. Primitive conditions for asymptotic
linearity can be quite strong and complicated. For example, it is known that
asymptotic linearity of estimators with a nonparametric first step often
requires some degree of smoothness in the functions being estimated, see
Ritov and Bickel (1990). Our purpose here is to bypass those conditions in
order to justify the Gateaux derivative formula for the influence function.
The formula for the influence function can then be used in all the important
ways outlined in Section 2.

It is also common to bypass regularity conditions when calculating the
influence function or asymptotic variance of parametric estimators. There
are well known formulae that allow us to do this, such as Hansen (1982) for
GMM estimators. The Gateaux derivative limit provides such a formula for
semiparametric estimators. It provides an influence function formula that
will be valid "under sufficient regularity conditions" analogously to the
GMM formula for parametric estimators.

\bigskip 

\begin{center}
\textsc{Appendix B: The influence function of semiparametric m estimators}
\end{center}

\bigskip 

In this Appendix we give the general structure of the influence function for
a semiparametric M-estimator and show that the FSIF is zero for any first
step that maximizes the same objective function as does the parameter of
interest. A maximization (M) estimator satisfies 
\begin{equation*}
\hat{\theta}=\arg \max_{\theta \in B}\hat{Q}(\theta ),
\end{equation*}%
for a function $\hat{Q}(\theta )$ that depends on the data and parameters. M
estimators have long been studied. A more general type that is useful when $%
\hat{Q}(\theta )$ is not continuous has $\hat{Q}(\theta )\geq \sup_{\theta
\in B}\hat{Q}(\theta )-\hat{R},$ where the remainder $\hat{R}$ is small in
large samples. The plin $\theta (F)$ of $\hat{\theta}$ will be the maximizer
of the probability limit of $\hat{Q}(\theta )$ under standard regularity
conditions. Thus, the influence function will depend only on the limit of
the objective function and so is not affected by whether $\hat{\theta}$ is
an approximate or exact maximizer of $\hat{Q}(\theta )$. The way we give of
calculating the influence function will work for many estimators of this
form, including those maximizing U-processes as considered by Sherman (1993).

We can use the Gateaux derivative to characterize the influence function for
semiparametric M-estimators. Let $Q_{\tau }(\theta )$ denote the plim of the
objective function $\hat{Q}(\theta )$ when the CDF of $W_{i}$ is $F_{\tau }$%
. Then under standard regularity conditions the plim of $\hat{\theta}$ is 
\begin{equation*}
\theta _{\tau }=\arg \max_{\theta \in \Theta }Q_{\tau }(\theta ).
\end{equation*}%
Suppose that $Q_{\tau }(\theta )$ is twice continuously differentiable in $%
\theta $ and $\theta _{\tau }$ is in the interior of the parameter set. Then 
$\theta _{\tau }$ satisfies the first order conditions $dQ_{\tau }(\theta
_{\tau })/d\theta =0$. By the implicit function theorem, for $\Lambda
=\partial ^{2}Q(\theta _{0})/\partial \theta \partial \theta ^{\prime }$ we
have%
\begin{equation*}
\frac{d\theta _{\tau }}{d\tau }=\left. -\Lambda ^{-1}\frac{\partial
^{2}Q_{\tau }(\theta _{0})}{\partial \tau \partial \theta }\right\vert
_{\tau =0}=-\Lambda ^{-1}\frac{\partial }{\partial \tau }\left\{ \frac{%
\partial Q_{\tau }(\theta _{0})}{\partial \theta }\right\} .
\end{equation*}%
Comparing this equation with equation (\ref{Inf deriv}) we see that the
influence function $\psi (w)$ of a semiparametric M estimator can be
calculated by evaluating the derivative with respect to $\tau $ of $dQ_{\tau
}(\theta _{0})/d\theta $ at the distribution $H_{w}^{\infty }$ with $W=w$
and premultiplying by $-\Lambda ^{-1}.$ For $\xi (W)$ such that $dQ_{\tau
}(\theta _{0})/d\theta =\int \xi (w)H(dw)$ the influence function of $\hat{%
\theta}$ will be%
\begin{equation*}
\psi (W)=-\Lambda ^{-1}\xi (W).
\end{equation*}%
This formula generalizes that of Newey (1994) for semiparametric GMM to
M-estimation.

For M-estimators, certain nonparametric components of $\hat{Q}(\theta )$ can
be ignored in deriving the influence function. The ignorable components are
those that have been \textquotedblleft concentrated out,\textquotedblright\
meaning they have a plim that maximizes the plim of $\hat{Q}(\theta )$. In
such cases the dependence of these functions on $\theta $ captures the whole
asymptotic effect of their estimation. To show this result, suppose that
there is a function $\gamma $ that depends on $\theta $ and possibly other
functions and a function $\tilde{Q}_{\tau }(\theta ,\gamma )$ such that $%
Q_{\tau }(\theta )=\tilde{Q}_{\tau }(\theta ,\gamma _{\tau })$ where 
\begin{equation*}
\gamma _{\tau }=\arg \max_{\gamma }\tilde{Q}_{\tau }(\theta ,\gamma ).
\end{equation*}%
Here $\tilde{Q}_{\tau }(\theta ,\gamma _{\tau })$ is the plim of $\hat{Q}%
(\theta )$ and $\gamma _{\tau }$ the plim of a nonparametric estimator on
which $\hat{Q}(\theta )$ depends, when $W$ has CDF $F_{\tau }$. Since $%
\gamma _{\tau }$ maximizes over all $\gamma $ it must maximize over $\tilde{%
\tau}$ as the function $\gamma _{\tilde{\tau}}$ varies. The first order
condition for maximization over $\tilde{\tau}$ is%
\begin{equation*}
\left. \frac{d\tilde{Q}_{\tau }(\theta ,\gamma _{\tilde{\tau}})}{d\tilde{\tau%
}}\right\vert _{\tilde{\tau}=\tau }=0.
\end{equation*}%
This equation holds identically in $\theta $, so that we can differentiate
both sides of the equality with respect to $\theta ,$ evaluate at $\theta
=\theta _{0}$ and $\tau =0,$ and interchange the order of differentiation to
obtain%
\begin{equation*}
\frac{\partial ^{2}\tilde{Q}(\theta _{0},\gamma _{\tau })}{\partial \tau
\partial \theta }=0.
\end{equation*}%
Then it follows by the chain rule that%
\begin{equation}
\frac{\partial ^{2}\tilde{Q}_{\tau }(\theta _{0},\gamma _{\tau })}{\partial
\tau \partial \theta }=\frac{\partial ^{2}\tilde{Q}_{\tau }(\theta
_{0},\gamma _{0})}{\partial \tau \partial \theta }+\frac{\partial ^{2}\tilde{%
Q}(\theta _{0},\gamma _{\tau })}{\partial \tau \partial \theta }=\frac{%
\partial ^{2}\tilde{Q}_{\tau }(\theta _{0},\gamma _{0})}{\partial \tau
\partial \theta }.  \label{concentrate}
\end{equation}%
That is, the influence function can be obtained by treating the limit $%
\gamma _{\tau }$ as if it were equal to the true value $\gamma _{0}$.

Equation (\ref{concentrate}) generalizes Proposition 2 of Newey (1994) and
Theorem 3.4 of Ichimura and Lee (2010) to objective functions that are not
necessarily a sample average of a function of $\theta $ and $\gamma .$ There
are many important estimators included in this generalization. One of those
is NPIV where the residual includes both parametric and nonparametric
components. The result implies that estimation of the function of the
nonparametric component $\gamma $ can be ignored in calculating the
influence function of $\theta .$ Another interesting estimator is partially
linear regression with generated regressors. There the estimation of the
nonparametric component can also be ignored in deriving the influence
function, just as in Robinson (1988), though the presence of generated
regressors will often affect the influence function, as in Hahn and Ridder
(2013, 2016) and Mammen, Rothe, and Schienle (2012).\pagebreak

\bigskip

\textsc{Appendix C: Endogenous orthogonality conditions with
misspecification.}

\bigskip

In this Appendix we derive the FSIF for endogenous orthgonality conditions
under overidentification and misspecification where%
\begin{equation*}
\bar{\pi}(X)=\pi (\rho (W,\gamma _{0})|X)\neq 0.
\end{equation*}%
The first order conditions for $\gamma _{\tau }=\arg \min_{\gamma }E_{\tau
}[\pi _{\tau }(\rho (W,\gamma _{\tau })|X)^{2}]$ give%
\begin{eqnarray*}
0 &=&E_{\tau }[\pi _{\tau }(\rho (W,\gamma _{\tau })|X)\pi _{\tau }(v_{\rho
\tau }(W)\Delta (Z)|X)] \\
&=&E_{\tau }[\pi _{\tau }(\rho (W,\gamma _{\tau })|X)v_{\rho \tau }(W)\Delta
(Z)]\text{ for all }\Delta \in \Gamma ,
\end{eqnarray*}%
identically in $\tau $. Define $\alpha (X,\Delta ):=$ $\pi (v_{\rho
}(W)\Delta (Z)|X)$ for $\Delta \in \Gamma $. Differentiating the previous
identity with respect to $\tau $ gives for all $\Delta \in \Gamma $%
\begin{eqnarray*}
0 &=&\frac{\partial }{\partial \tau }E[\pi _{\tau }(\rho (W,\gamma _{\tau
})|X)\alpha (X,\Delta )]+\int \phi _{1}(w,\Delta )H(dw)+T_{v_{\rho }}(\Delta
), \\
\phi _{1}(w,\Delta ) &:&=\bar{\pi}(X)v_{\rho }(W)\Delta (Z)-E[\bar{\pi}%
(X)v_{\rho }(W)\Delta (Z)],\text{ }T_{v_{\rho }}(\Delta ):=\text{ }\frac{%
\partial }{\partial \tau }E[\bar{\pi}(X)v_{\rho \tau }(W)\Delta (W)].
\end{eqnarray*}%
where $v_{\rho }(W)=v_{\rho 0}(W).$ Solving gives%
\begin{equation}
\frac{\partial }{\partial \tau }E[\pi _{\tau }(\rho (W,\gamma _{\tau
})|X)\alpha (X,\Delta )]=-\int \phi _{1}(w,\Delta )H(dw)-T_{v_{\rho
}}(\Delta )  \label{mis der 1}
\end{equation}%
for all $\Delta \in \Gamma .$

Next we use the orthogonality condition for the projection that for all $%
b\in \mathcal{B}$ 
\begin{equation*}
E_{\tau }[\rho (W,\gamma _{\tau })b(X)]=E_{\tau }[\pi _{\tau }(\rho
(W,\gamma _{\tau })|X)b(X)].
\end{equation*}%
Because $\mathcal{A}$ is a subset of $\mathcal{B}$ it follows that 
\begin{equation*}
E_{\tau }[\rho (W,\gamma _{\tau })\alpha (X,\Delta )]=E_{\tau }[\pi _{\tau
}(\rho (W,\gamma _{\tau })|X)\alpha (X,\Delta )]\text{ for all }\Delta \in
\Gamma ,
\end{equation*}%
identically in $\tau .$ Differentiating both sides of this identify with
respect to $\tau $ and applying the chain rule gives%
\begin{eqnarray*}
\frac{\partial }{\partial \tau }E_{\tau }[\rho (W,\gamma _{\tau })\alpha
(X,\Delta )] &=&\frac{\partial }{\partial \tau }E_{\tau }[\bar{\pi}(X)\alpha
(X,\Delta )]+\frac{\partial }{\partial \tau }E[\pi _{\tau }(\rho (W,\gamma
_{\tau })|X)\alpha (X,\Delta )] \\
&=&\frac{\partial }{\partial \tau }E_{\tau }[\bar{\pi}(X)\alpha (X,\Delta
)]-\int \phi _{1}(w,\Delta )H(dw)-T_{v_{\rho }}(\Delta ) \\
&=&-\int \phi _{\Gamma }(w,\Delta )H(dw)-T_{v_{\rho }}(\Delta ), \\
\phi _{\Gamma }(w,\Delta ) &=&\bar{\pi}(X)\{v_{\rho }(X)\Delta (Z)-\alpha
(X,\Delta )\},
\end{eqnarray*}%
for all $\Delta \in \Gamma $ where the second equality follows by equation (%
\ref{mis der 1}) and the third equality equality follows by $E[\bar{\pi}%
(X)\alpha (X,\Delta )]=E[\bar{\pi}(X)v_{\rho }(W)\Delta (Z)].$ Applying the
chain rule to the left-hand side and solving then gives%
\begin{eqnarray}
-\frac{\partial }{\partial \tau }E[\rho (W,\gamma _{\tau })\alpha (X,\Delta
)] &=&\frac{\partial }{\partial \tau }E_{\tau }[\rho (W,\gamma _{0})\alpha
(X,\Delta )]+\int \phi _{\Gamma }(w,\Delta )H(dw)+T_{v_{\rho }}(\Delta )
\label{mis inf func} \\
&=&\int \{\rho (w,\gamma _{0})\alpha (x,\Delta )+\phi _{\Gamma }(w,\Delta
)\}H(dw)+T_{v_{\rho }}(\Delta )\text{,}  \notag
\end{eqnarray}%
for all $\Delta \in \Gamma ,$ where the last equality follows by the first
order condition at $\tau =0$ that implies $E[\rho (W,\gamma _{0})\alpha
(X,\Delta )]=0$ for all $\Delta .$ Suppose that there exists $b_{m}$ such
that the projection of $b_{m}$ on $\mathcal{A}$ is $\alpha (X,\Delta _{m})$
for some $\Delta _{m}\in \Gamma $ and%
\begin{equation*}
\Pi (v_{m}(Z)|Z)=-\Pi (v_{\rho }(W)b_{m}(X)|Z).
\end{equation*}%
Then by $\gamma _{\tau }(Z)\in \Gamma ,$%
\begin{eqnarray}
E[v_{m}(Z)\gamma _{\tau }(Z)] &=&E[\Pi (v_{m}(Z)|Z)\gamma _{\tau
}(Z)]=-E[\Pi (v_{\rho }(W)b_{m}(X)|Z)\gamma _{\tau }(Z)]  \label{mis deri 2}
\\
&=&-E[v_{\rho }(W)b_{m}(X)\gamma _{\tau }(Z)]=-E[b_{m}(X)\pi (v_{\rho
}(W)\gamma _{\tau }(Z)|X)]  \notag \\
&=&-E[\alpha (X,\Delta _{m})\pi (v_{\rho }(W)\gamma _{\tau
}(Z)|X)]=-E[\alpha (X,\Delta _{m})v_{\rho }(W)\gamma _{\tau }(Z)].  \notag
\end{eqnarray}%
Then differentiating gives%
\begin{eqnarray*}
\frac{\partial }{\partial \tau }E[m(W,\gamma _{\tau })] &=&\frac{\partial }{%
\partial \tau }E[v_{m}(Z)\gamma _{\tau }(Z)]=-\frac{\partial }{\partial \tau 
}E[\alpha (X,\Delta _{m})v_{\rho }(W)\gamma _{\tau }(Z)] \\
&=&-\frac{\partial }{\partial \tau }E[\alpha (X,\Delta _{m})\rho (W,\gamma
_{\tau })] \\
&=&\int \{\rho (w,\gamma _{0})\alpha (x,\Delta _{m})+\phi _{\Gamma
}(w,\Delta _{m})\}H(dw)+T_{v_{\rho }}(\Delta _{m})
\end{eqnarray*}%
where the first equality follows by Assumption 3, the second equality by
equation (\ref{mis deri 2}), the third equality by Assumption 4, and the
fourth equality by equation (\ref{mis inf func}). Combining this last
equation with the conditions on which it depends gives the following result:

\bigskip

\textsc{Proposition C1:} \textit{If i) Assumptions 3-4 are satisfied; ii)
there exists }$b_{m}(X)$ and $\Delta _{m}\in \Gamma $ such that $\alpha
(X,\Delta _{m})$ is the projection of $b_{m}(X)$ on $\mathcal{A}$ and $\Pi
(v_{m}(Z)|Z)=\Pi (v_{\rho }(W)b_{m}(X)|Z);$ \textit{and iii) there is }$\phi
_{\rho }(w)$\textit{\ such that} $\partial E[\bar{\pi}(X)v_{\rho \tau
}(W)\Delta _{m}(W)]/\partial \tau =\int \phi _{\rho }(w)H(dw)$ \textit{then
the FSIF is}%
\begin{equation*}
\phi (w,\gamma ,\alpha )=\alpha (x,\Delta _{m})\rho (w,\gamma )+\bar{\pi}%
(x)\{v_{\rho }(x)\Delta _{m}(z)-\pi (v_{\rho }(X)\Delta _{m}(Z)|X=x)\}+\phi
_{\rho }(w).
\end{equation*}

\bigskip

This expression for the influence function contains the term $\phi _{\rho
}(w)$ which is the influence function of $E[\bar{\pi}(X)v_{\rho \tau
}(W)\Delta _{m}(Z)].$ This $\phi _{\rho }(w)$ need not exist. In particular
for quantile orthogonality conditions where $v_{\rho \tau }(W)$ depends on
the conditional pdf of $Y$ given $Z$ and $X$ evaluated at the point $%
Y=\gamma _{0}(Z)$ it seems that this $\phi _{\rho }(w)$ generally does not
exist. In that case the NPIV estimator will not root-n consistent under
misspecification. This problem does not appear to be present for expectiles,
where $E[\bar{\pi}(X)v_{\rho \tau }(W)\Delta _{m}(Z)]$ can be shown to have
an influence function.

Ai and Chen (2007, p. 40) gave an influence function for a function of the
solution to a conditional moment restriction under misspecification. In this
case the expression given in Proposition 3 is analogous to that in Ai and
Chen (2007). Proposition C1 generalizes that expression to orthogonality
conditions.

\bigskip 

\begin{center}
\textsc{References}
\end{center}

\bigskip 

Ackerberg, D., X. Chen, J. Hahn (2012): "A Practical Asymptotic Variance
Estimator for Two-Step Semiparametric Estimators," \textit{The Review of
Economics and Statistics} 94, 481-498.

Ackerberg, D., X. Chen, J. Hahn and Z. Liao (2014): \textquotedblleft
Asymptotic Efficiency of Semiparametric Two-step GMM\textquotedblright\ 
\textit{Review of Economic Studies} 81, 919-943.

Ai, C. and X. Chen (2003): ``Efficient Estimation of Models with Conditional
Moment Restrictions Containing Unknown Functions," \textit{Econometrica }71,
1795--1843.

Ai, C. and X. Chen (2007): ``Estimation of Possibly Misspecified
Semiparametric Conditional Moment Restriction Models with Different
Conditioning Variables," \textit{Journal of Econometrics} 141, 5--43.

Ai, C. and X. Chen (2012): ``The Semiparametric Efficiency Bound for Models
of Sequential Moment Restrictions Containing Unknown Functions," \textit{%
Journal of Econometrics} 170, 442--457.

Albrecht, J., A. Bjorklund, and S. Vroman (2003): "Is There a Glass Ceiling
in Sweden?" \textit{Journal of Labor Economics} 21, 145-177.

Amemiya, T. (1985): \textit{Advanced Econometrics}, Cambridge, Harvard
University Press.

Andrews, D. W. K. (1994): \textquotedblleft Asymptotics for Semiparametric
Econometric Models via Stochastic Equicontinuity,\textquotedblright\ \textit{%
Econometrica} 62, 43--72.

Andrews, D.W.K. (2011): ``Examples of L2-Complete and Boundedly-Complete
Distributions,\textquotedblright\ Cowles Foundation Discussion Paper No.
1801, Yale University.

Andrews, I., M. Gentzkow, and J.M. Shapiro (2017): "Measuring the
Sensitivity of Parameter Estimates to Estimation of Moments," \textit{%
Quarterly Journal of Economics,} 132, 1553-1592.

Angrist, J.D. and A.B. Krueger (1991): "Does Compulsory School Attendance
Affect Schooling and Earnings?" \textit{The Quarterly Journal of Economics}
106, 979-1014.

Bajari, P., V. Chernozhukov, H. Hong, and D. Nekipelov (2009):
"Nonparametric and Semiparametric Analysis of a Dynamic Discrete Game,"
working paper, Stanford.

Bajari, P., H. Hong, J. Krainer, and D. Nekipelov (2010): \textquotedblleft
Estimating Static Models of Strategic Interactions," \textit{Journal of
Business and Economic Statistics} 28, 469-482.

Bickel, P, C. Klaasen, Y. Ritov, and J. Wellner (1993): \textit{Efficient
and Adaptive Estimation for Semiparametric Models, }Washington, Johns
Hopkins.

Blundell, R. and R. Matzkin (2014): "Control Functions In Nonseparable
Simultaneous Equations Models," \textit{Quantitative Economics} 5, 271--295.

Blundell, R., J. Horowitz, and M.\ Parey (2017): Nonparametric Estimation of
a Nonseparable Demand Function Under the Slutzky Inequality Restriction," 
\textit{The Review of Economics and Statistics} 99: 291--304.

Carone, M., A.R. Luedtke, M.J. van der Laan (2016): \textquotedblleft Toward
computerized efficient estimation in infinite dimensional models," arXiv:
1608.08717v1 31Aug2016.

Chaudhuri, P.K. Doksum, and A. Samarov (1997): "On Average Derivative
Quantile Regression," \textit{The Annals of Statistics} 25, 715-744.

Chen, X., O. Linton, and I. van Keilegom, (2003): \textquotedblleft
Estimation of Semiparametric Models When the Criterion Function is not
Smooth,\textquotedblright\ \textit{Econometrica} 71, 1591--1608.

Chen, X., and Z. Liao (2015): \textquotedblleft Sieve Semiparametric GMM
Under Weak Dependence," \textit{Journal of Econometrics} 189, 163-186.

Chen, X. and D. Pouzo (2015): "Sieve Wald and QLR Inferences on
Semi/Nonparametric Conditional Moment Models," \textit{Econometrica} 83,
1013-1079.

Chen, X. and A. Santos (2015): "Overidentification in Regular Models," 
\textit{Econometrica} 86, 1771-1817.

Chernozhukov, V., I. Fernandez-Val, and B. Melly (2013): "Inference on
Counterfactual Distributions," \textit{Econometrica }81, 2205-2268.

Chernozhukov, V., D. Chetverikov, M. Demirer, E. Duflo, C. Hansen, W. Newey,
and J. Robins (2018): "Double/Debiased Machine Learning for Treatment and
Structural Parameters," \textit{Econometrics Journal }21, C1--C68.

Chernozhukov, V., J.-C. Escanciano, H. Ichimura, W. Newey, and J. Robins
(2020): \textquotedblleft Locally Robust Semiparametric Estimation,"
https://arxiv.org/pdf/1608.00033.pdf.

Conley, T.J., C.B. Hansen, and P.E. Rossi (2012): "Plausibly Exogenous," 
\textit{The Review of Economics and Statistics} 94, 260-272.

Darolles, S., Y. Fan, J. P. Florens, and E. Renault (2011):
\textquotedblleft Nonparametric Instrumental Regression," \textit{%
Econometrica} 79, 1541--1565.

Firpo, S., N.M. Fortin, and T. Lemieux (2009): "Unconditional Quantile
Regressions,"\textit{\ Econometrica} 77, 953--973.

Gentzkow, M., and J.M. Shapiro (2015): "Measuring the Sensitivity of
Parameter Estimates to Sample Statistics," working paper.

Goldstein, L. and K. Messer (1992): ``Optimal Plug-in Estimators for
Nonparametric Functional Estimation,'' \textit{Annals of Statistics} 20,
1306--1328.

Hahn, J., (1998): ``On the Role of Propensity Score in Efficient
Semiparametric Estimation of Average Treatment Effects" \textit{Econometrica 
}66, 315-332.

Hahn, J. and G. Ridder (2013): ``The Asymptotic Variance of Semi-parametric
Estimators with Generated Regressors," \textit{Econometrica} 81, 315-340.

Hahn, J. and G. Ridder (2016): ``Three-stage Semi-Parametric Inference:
Control Variables and Differentiability," working paper.

Hampel, F. R. (1974): \textquotedblleft The Influence Curve and Its Role In
Robust Estimation,\textquotedblright\ \textit{Journal of the American
Statistical Association} 69, 383--393.

Hansen, L.P. (1982): \textquotedblleft Large Sample Properties of
Generalized Method of Moments Estimators," \textit{Econometrica} 50,
1029-1054.

Hausman, J.A. (1978): "Specification Tests in Econometrics," \textit{%
Econometrica} 46, 1251-1271.

Hausman, J. A. and W. K. Newey (2016): \textquotedblleft Individual
Heterogeneity and Average Welfare," \textit{Econometrica} 84, 1225-1248.

Hausman, J.A. and W.K. Newey (2017): \textquotedblleft Nonparametric Welfare
Analysis," \textit{Annual Review of Economics }9, 521--546.

Hirano, K., G.W. Imbens, G. Ridder (2003): \textquotedblleft Efficient
Estimation of Average Treatment Effects Using the Propensity Score," \textit{%
Econometrica} 71, 1161-1189.

Huber, P. (1981): \textit{Robust Statistics,}\ New York: John Wiley and Sons.

Ichimura, H. (1993): ``Semiparametric Least Squares (SLS) and Weighted SLS
Estimation of Single-index Models," \textit{Journal of Econometrics} 58,
71-120.

Ichimura, H. and S. Lee (2010): \textquotedblleft Characterization of the
asymptotic distribution of semiparametric M-estimators,\textquotedblright\ 
\textit{Journal of Econometrics} 159, 252--266.

Imbens, G.W. (1997): "One-Step Estimators for Over-Identified Generalized
Method of Moments Models," \textit{The Review of Economic Studies} 64,
359--383.

Klein, R.W. and R.H. Spady (1993): \textquotedblleft An Efficient
Semiparametric Estimator for Binary Response Models," \textit{Econometrica}
61, 387-421.

Koenker, R. and G. Bassett (1978): "Regression Quantiles," \textit{%
Econometrica} 46, 33-50.

Kress, R. (1989):\textit{\ Linear Integral Equations, }New York:
Springer-Verlag.

Lebesgue, H. (1904): \textit{Le\c{c}ons sur l'Int\'{e}gration et la
Recherche des Fonctions Primitives}, Paris: Gauthier-Villars.

Luedtke, A.R., M. Carone, M.J. van der Laan (2015): \textquotedblleft Toward
computerized efficient estimation in infinite dimensional models," arXiv:
1608.08717.

Machada, J.A.F and J. Mata (2005): "Counterfactual Decomposition of Changes
in Wage Distributions Using Quantile Regression," \textit{Journal of Applied
Econometrics} 20, 445-465.

Mammen, E., C. Rothe, M. Schienle (2012): \textquotedblleft Nonparametric
Regression with Nonparametrically Generated Covariates,\textquotedblright\ 
\textit{Annals of Statistics} 40, 1132--1170.

Melly, B. (2005): "Decomposition of Differences in Distribution Using
Quantile Regression," \textit{Labour Economics} 12, 577-590.

Mukhin, Y. (2019): "Sensitivity of Regular Estimators," arXiv 1805.08883.

Nagar, A.L. (1959): "The Bias and Moment Matrix of the General k-Class
Estimators of the Parameters in Simultaneous Equations," \textit{Econometrica%
} 27, 575-595.

Newey, W.K. (1991): \textquotedblright Uniform Convergence in Probability
and Stochastic Equicontinuity,\textquotedblright\ \textit{Econometrica} 59,
1161-1167.

Newey, W.K. (1994): \textquotedblleft The Asymptotic Variance of
Semiparametric Estimators,\textquotedblright\ \textit{Econometrica} 62,
1349--1382.

Newey, W.K. and J.L. Powell (1987): "Asymmetric Least Squares Estimation and
Testing," \textit{Econometrica} 55, 819-847.

Newey, W.K., and J.L. Powell (1989): \textquotedblleft Instrumental Variable
Estimation of Nonparametric Models," presented at Econometric Society winter
meetings, 1988.

Newey, W.K., and J.L. Powell (2003): \textquotedblleft Instrumental Variable
Estimation of Nonparametric Models," \textit{Econometrica} 71, 1565-1578.

Powell, J.L., J.H. Stock, and T.M. Stoker (1989): \textquotedblleft
Semiparametric Estimation of Index Coefficients," \textit{Econometrica} 57,
1403-1430.

Pratt, J.W. (1981): "Concavity of the Log Likelihood," \textit{Journal of
the American Statistical Association} 76, 103-106.

Ritov, Y. and P.J. Bickel (1990): \textquotedblleft Achieving Information
Bounds in Non and Semiparametric Models," \textit{Annals of Statistics }18,
925-938.

Robinson, P.M. (1988): \textquotedblleft Root-N-Consistent Semiparametric
Regression," \textit{Econometrica} 56, 931-954.

Severini, T. and G. Tripathi (2012): ``Efficiency Bounds for Estimating
Linear Functionals of Nonparametric Regression Models with Endogenous
Regressors," \textit{Journal of Econometrics} 170, 491-498.

Sherman, R. (1993): \textquotedblleft The Limiting Distribution of the
Maximum Rank Correlation Estimator," \textit{Econometrica} 61, 123-137.

Stein, C. (1956): "Efficient Nonparametric Testing and Estimation," in 
\textit{Proceedings of the Third Berkeley Symposium in Mathematical
Statistics and Probability, vol. 1, }187-196, Berkeley, University of
California Press.

Van der Vaart, A.W. (1991): \textquotedblleft On Differentiable
Functionals,\textquotedblright\ \textit{Annals of Statistics} 19, 178--204.

Van der Vaart, A.W. (1998): "Asymptotic Statistics," Cambridge: Cambridge
University Press.

Van der Vaart, A. W. and J. A. Wellner (1996): \textit{Weak Convergence and
Empirical Processes}, New York: Springer-Verlag.

Von Mises, R. (1947): \textquotedblleft On the Asymptotic Distribution of
Differentiable Statistical Functions," \textit{Annals of Mathematical
Statistics} 18, 309-348.

Wheeden, R.L. and A. Zygmund (1977): \textit{Measure and Integral -- An
introduction to Real Analysis}, Marcel Dekker.

\end{document}